\documentclass[11pt, onecolumn]{IEEEtran}
\def \Paren#1{{\left({#1}\right)}}
\def \absv#1{{|#1|}}
\newcommand{\expectation}[1]{\mathbb{E}\left[#1\right]}
\def \binomial{\text{Binom}}

\usepackage[utf8]{inputenc} 
\usepackage[T1]{fontenc}    
\usepackage{hyperref}       
\usepackage{url}            
\usepackage{booktabs}       
\usepackage{nicefrac}       
\usepackage{microtype}      
\usepackage{bbm}
\usepackage{graphicx}
\usepackage{color}
\usepackage[dvipsnames]{xcolor}
\usepackage[T1]{fontenc}
\usepackage{subcaption}
\usepackage{dsfont}
\usepackage{amsfonts}
\usepackage{graphicx,wrapfig}
\usepackage[T1]{fontenc}
\usepackage{amsmath}
\usepackage{mathtools, cuted}
\usepackage{amsthm}
\usepackage{amstext}
\usepackage{amssymb}
\usepackage{mathrsfs}
\usepackage[square,numbers, sort&compress]{natbib}
\usepackage{tikz}
\usepackage{marginnote}
\usepackage{tkz-tab}
\usetikzlibrary{topaths,calc}
\usepackage{pgfplots}
\usepackage{stackengine}
\usepackage{algorithmic}
\usepackage[ruled]{algorithm} 

\DeclareMathAlphabet{\mathbsf}{OT1}{cmss}{bx}{n}
\DeclareMathAlphabet{\mathssf}{OT1}{cmss}{m}{sl}
\DeclareSymbolFont{bsfletters}{OT1}{cmss}{bx}{n}  
\DeclareSymbolFont{ssfletters}{OT1}{cmss}{m}{n}

\usetikzlibrary{arrows,positioning, shapes}
\usepackage{float}

\usepackage{setspace}
\usepackage{hyperref}
\hypersetup{colorlinks,
   colorlinks,
    citecolor=red,
    filecolor=green,
  linkcolor=blue,
    linktoc=page,
   urlcolor=blue
}

\hypersetup{
    colorlinks,
    linkcolor={blue},
    citecolor={green!60!black},
    urlcolor={blue}
}

\def\R{\mathbb{R}}

\def\argmin{\mathop{\rm arg\, min}}
\def\eps{\varepsilon}

\def\E{{\mathbb E}}
\def\F{{\mathcal F}}
\def\H{{\mathcal H}}

\def\P{{\mathcal P}}
\def\Q{{\mathcal Q}}

\def\X{{\mathcal X}}

\def\N{{\mathcal N}}

\def\Z{{\mathcal Z}}

\def\V{{\mathcal V}}

\def\sBer{{\mathsf{Bernoulli}}}

\def\sK{{\mathsf K}}

\def\sEs{{\mathsf E_{e^{\varepsilon}}}}
\def\sE{{\mathsf E}}

\def \var {{\mathsf {var}   }}

\def\tv{{\mathsf {TV}}}
\def\kl{{\mathsf {KL}}}

\usepackage[font=small,labelsep=space]{caption}
\DeclareCaptionLabelSeparator{dot}{.~}
\newcounter{example}
\newenvironment{example}[1][]{\refstepcounter{example}\par\medskip
   \noindent \textit{Example~\theexample. #1} \rmfamily}{\medskip}

\newtheorem{theorem}{Theorem}
\newtheorem{corollary}{Corollary}

\newtheorem{proposition}{Proposition}
\newtheorem{lemma}{Lemma}
\newtheorem{remark}{Remark}

\usepgfplotslibrary{fillbetween}
\usepackage{times}
\usepackage{tikz}
\usepackage{amsmath, bm}
\usepackage{verbatim}
\usetikzlibrary{arrows,shapes}
\tikzstyle{RectObject}=[rectangle,fill=white,draw,line width=0.2mm]
\tikzstyle{line}=[draw]
\tikzstyle{arrow}=[draw, -latex]
\usetikzlibrary{decorations.pathmorphing}

\definecolor{DukeBlue}{HTML}{001A57}
\definecolor{DarkRed}{rgb}{0.75, 0.0, 0.0}
\definecolor{DarkGreen}{rgb}{0.0, 0.5, 0.0}

\usepackage[normalem]{ulem}

\def\dist{\alpha}
\newcommand{\newhz}[1]{\textcolor{black}{#1}}

\allowdisplaybreaks
\usepackage[letterpaper, left=1in, right=1in, bottom=1in, top=1in]{geometry}
\usepackage{setspace}
\date{}

\begin{document}

	\title{\vspace{5.5mm}Contraction of Locally Differentially Private Mechanisms}
\author{Shahab Asoodeh\thanks{S. Asoodeh is with the Department of Computing and Software, McMaster University, Hamilton, ON L8S 1C7, Canada.   Email: \texttt{asoodeh@mcmaster.ca}.  Much of this work was completed while S.A.\ was a visiting research scientist at the Meta's Statistics and Privacy Team.} and Huanyu Zhang\thanks{H. Zhang is with Meta Platforms, Inc., New York, NY 10003, USA. Email: \texttt{huanyuzhang@meta.com}.}} 
	
	\maketitle
	\begin{abstract}
  We investigate the contraction properties of locally differentially private mechanisms. More specifically, we derive tight upper bounds on the divergence between $P\sK$ and $Q\sK$ output distributions of an $\eps$-LDP mechanism $\sK$ in terms of a divergence between the corresponding input distributions $P$ and $Q$, respectively. Our first main technical result presents a sharp upper bound on the $\chi^2$-divergence $\chi^2(P\sK\|Q\sK)$ in terms of $\chi^2(P\|Q)$ and $\eps$.  We also show that the same result holds for a large family of divergences, including KL-divergence and squared Hellinger distance. The second main technical result gives an upper bound on $\chi^2(P\sK\|Q\sK)$ in terms of total variation distance $\tv(P, Q)$ and $\eps$. We then utilize these bounds to establish locally private versions of the van Trees inequality, Le Cam's, Assouad's, and the mutual information methods ---powerful tools for bounding minimax estimation risks. These results are shown to lead to tighter privacy analyses than the state-of-the-arts in several statistical problems such as entropy and discrete distribution estimation, non-parametric density estimation, and hypothesis testing. 
\end{abstract}



\section{Introduction}
Local differential privacy (LDP)  has now become a standard definition for individual-level privacy in machine learning. Intuitively, a randomized mechanism (i.e., a channel) is said to be locally differentially private if its output does not vary significantly with arbitrary perturbation of the input. More precisely, a mechanism is $\eps$-LDP if the privacy loss random variable, defined as the log-likelihood ratio of the output for any two different inputs, is smaller than $\eps$.

Since its formal introduction  \cite{evfimievski2003limiting, Shiva_subsampling}, LDP has been extensively incorporated into statistical problems, e.g., locally private mean estimation problem  \cite{Duchi_LDP_MinimaxRates,Duchi_Federatedprotection, Duchi_FisherInfo, Duchi_interactivity,Shuffled_FL, Chen_Breaking, LDP_gaboardi19a, LDP_Acharya1, Asoodeh_LDP_Equivalent, Optimal_mean_LDP, Ampli_Shuffling, cp_SGD,VQ_SGD,rohde2020,Jayadev_Unified, Jayadev_chiSquare_Contraction},  and  locally private distribution estimation problem \cite{LDP_DistributionEstimation, Disribution_estimation_hadamard, kairouz16_LDPEstimation, LDP_Fisher, Jayadev_Unified, Optimal_compressionLDP_Feldman, Optimal_compressionLDP_Kairouz, feldman2022private}. 
The fundamental limits of such statistical problems under LDP are typically characterized using information-theoretic frameworks such as Le Cam's, Assouad's, and Fano's methods \cite{Yu1997}. A critical building block for sharp privacy analysis in such methods turns out to be the \textit{contraction coefficient} of LDP mechanisms. Contraction coefficient $\eta_f(\sK)$ of a mechanism $\sK$ under an $f$-divergence is a quantification of how much the data processing inequality can be strengthened: It is the smallest $\eta\leq 1$ such that $D_f(P\sK\|Q\sK)\leq \eta D_f(P\|Q)$ for any distributions $P$ and $Q$, where $P\sK$ denotes the output distribution of $\sK$ when its input is sampled from $P$. 

Studying statistical problems under local privacy through the lens of contraction coefficients was initiated by Duchi et al. \citep{Duchi_LDP_MinimaxRates, Duchi2016MinimaxOP} in which sharp minimax risks for locally private mean estimation problems were characterized for sufficiently small $\eps$. As the main technical result, they showed that any $\eps$-LDP mechanism $\sK$ satisfies  
\begin{equation}\label{Duchi1_KL}
    D_\kl(P\sK\|Q\sK) \leq \min\{4, e^{2\eps}\} (e^\eps-1)^2\tv^2(P, Q),
\end{equation}
where $D_\kl(\cdot\|\cdot)$ and $\tv(\cdot, \cdot)$ denote KL-divergence and total variation distance, respectively. In light of the Pinsker's inequality $2\tv^2(P, Q)\leq D_\kl(P\|Q)$, this result gives an upper bound on $\eta_\kl(\sK)$ the contraction coefficient under KL-divergence. However, thanks to the data processing inequality, this bound becomes vacuous if the coefficient in \eqref{Duchi1_KL} is strictly bigger than $1$ (i.e., $\eps$ is not sufficiently small). More recently, Duchi and Ruan \citep[Proposition 8]{Duchi_FisherInfo} showed a similar upper bound for $\chi^2$-divergence:
\begin{equation}\label{Duchi3_tv}
    \chi^2(P\sK\|Q\sK)\leq 4 (e^{\eps^2}-1)\tv^2(P,Q).
\end{equation}
According to Jensen's inequality $4\tv^2(P, Q)\leq \chi^2(P\| Q)$, and thus \eqref{Duchi3_tv} implies an upper bound on $\eta_{\chi^2}(\sK)$ the contraction coefficient under $\chi^2$-divergence. Analogously, this bound is non-trivial only for sufficiently small $\eps$.   
Similar upper bounds on the contraction coefficients under total variation distance and hockey-stick divergence were determined in \cite{kairouz2014extremal_JMLR} and \cite{Asoodeh_LDP_Equivalent}, respectively. 
Results of this nature are recurrent themes in privacy analysis in statistics and machine learning, see \cite{LDP_Acharya1, Jayadev_Unified, Jayadev_role_interactivity, Acharya_discrete_estimationLDP} for other examples of such results. 

In this work, we develop a framework for characterizing tight upper bounds on $D_\kl(P\sK\|Q\sK)$ and $\chi^2(P\sK\|Q\sK)$ for any LDP mechanisms. We achieve this goal via two different approaches: \textit{(i)} indirectly by bounding  $\eta_{\kl}(\sK)$ and $\eta_{\chi^2}(\sK)$, and \textit{(ii)} directly by deriving inequalities of the form \eqref{Duchi1_KL} and \eqref{Duchi3_tv} that are considerably tighter for all $\eps\geq 0$. In particular, our main contributions are:
\begin{enumerate}
    \item We obtain a sharp upper bound on $\eta_{\chi^2}(\sK)$ for any $\eps$-LDP mechanism $\sK$ in Theorem~\ref{thm_chiSDPI}, and show that this bound holds for a large family of divergences, including KL-divergence and squared Hellinger distance. 
    \item We derive upper bounds for $D_\kl(P\sK\|Q\sK)$ and $\chi^2(P\sK\|Q\sK)$ in terms of $\tv(P, Q)$ and the privacy parameter $\eps$ in Theorem~\ref{Thm:Chi_TV}. While upper bounds in \eqref{Duchi1_KL} and \eqref{Duchi3_tv} scale as $O(e^{2\eps})$ and $O(e^{\eps^2})$, respectively, ours scales as $O(e^\eps)$, thus significantly improving over those bounds for practical range of $\eps$ (that is $\eps\geq \frac{1}{2}$).  
    \item We use our main results to develop a systemic framework for quantifying the cost of local privacy in several statistical problems under the ``sequentially interactive'' setting. Our framework enables us to improve and generalize several existing results, and also produce new results beyond the reach of existing techniques. In particular, we study the following problems:
    \paragraph{\textbf{Locally private Fisher information}}
    We show that the Fisher information matrix $I_{Z^n}(\theta)$ of parameter $\theta$ given a privatized sequence $Z^n\coloneqq (Z_1, \dots, Z_n)$ of $X^n\stackrel{\text{iid}}{\sim} P_\theta$ satisfies $I_{Z^n}(\theta)\preccurlyeq n\big[\frac{e^{\eps}-1}{e^{\eps}+1}\big]^2 I_X(\theta)$ (Lemma~\ref{Lemm_FI_chi}). This result then directly leads to a private version of 
    the van Trees inequality (Corollary~\ref{cor:PrivatevanTrees}) that is a classical approach for lower bounding the minimax quadratic risk. In Appendix~\ref{app:vanTrees}, we also provide a private version of the Cram\'er-Rao bound, provided that there exist unbiased private estimators.  It is worth noting that Barnes et al. \cite{LDP_Fisher} recently investigated locally private Fisher information under certain assumptions regarding the regularity of $P_\theta$. More specifically, they derived various upper bounds on $\mathsf{Tr}(I_{Z}(\theta))$ for $\eps\geq 0$, when $\log f_\theta(X)$ is either sub-exponential or sub-Gaussian, or when $\E[(u^{\mathsf{T}}\nabla \log f_\theta(X))^2]$ is bounded for any unit vector $u\in \R^d$, where $f_\theta$ is the density of $P_\theta$ with respect to the Lebesgue measure. 
 In contrast, Lemma~\ref{Lemm_FI_chi} establishes a similar upper bound for small $\eps$ (i.e., $\eps\in [0,1]$) but without imposing  such regularity conditions.
     
    %
    
    \paragraph{\textbf{Locally private Le Cam's and Assouad's methods}}  Following \cite{Duchi_LDP_MinimaxRates}, we establish locally private versions of  
    Le Cam's and Assouad's methods \cite{lecam1973, Yu1997} that are demonstrably stronger than those presented in \cite{Duchi_LDP_MinimaxRates} (Theorems~\ref{thm:private_le_cam} and \ref{thm:private_assouad}). 
    We then used our private Le Cam's method to study the problem of entropy estimation under LDP where the underlying distribution is known to be supported over $\{1, \dots, k\}$ (Corollary~\ref{cor:entropt_estimation}).    
    As applications of our private Assouad's method, we study two problems. First, 
    we derive a lower bound for $\ell_h$ minimax risk in the locally private distribution estimation problem which improves the constants of the state-of-the-art lower bounds~\cite{LDP_DistributionEstimation} in the special cases $h=1$ and $h=2$, and leads to the same order analysis for general $h\geq 1$  in \cite{Jayadev_Unified}. We also provide an upper bound by generalizing the Hadamard response~\cite{Disribution_estimation_hadamard} to $\ell_h$-norm with $h\ge2$ which matches the lower bound under some mild conditions.  
    Second, we study private non-parametric density estimation when the underlying density is assumed to be H\"older continuous and derive a lower bound for $\ell_h$ minimax risk in Corollary~\ref{cor:DensityEstimation}. Unlike the best existing result \cite{LDP_density_estimation}, our lower bound holds for all $\eps\geq 0$.
    

     \paragraph{\textbf{Locally private mutual information method}} Recently, mutual information method \citep[Section 11]{Yihong_Lecturenote_Stat} has been proposed as a more flexible information-theoretic technique for bounding the minimax risk. We invoke Theorem~\ref{thm_chiSDPI} to provide (for the first time) a locally private version of the mutual information bound in Theorem~\ref{thm:MIM}. To demonstrate the flexibility of this result, we consider the Gaussian
    location model where the goal is to privately estimate $\theta\in \Theta$ from $X^n\stackrel{\text{iid}}{\sim}\N(\theta, \sigma^2\mathnormal{I}_d)$. 
    Most existing results (e.g., \cite{Duchi_LDP_MinimaxRates, Duchi_FisherInfo, LDP_Fisher, Duchi_interactivity}) assume $\ell_2$-norm as the loss and unit $\ell_\infty$-ball or unit $\ell_2$-ball as $\Theta$. However, our result presented in Corollary~\ref{cor:GaussianLocation_general} holds for any \textit{arbitrary} loss functions and any \textit{arbitrary} set $\Theta$ (e.g., $\ell_h$-ball for any $h\geq 1$).

     \paragraph{\textbf{Locally private hypothesis testing}} Given $n$ i.i.d.\ samples and two distributions $P$ and $Q$, we derive upper and lower bounds for $\mathsf{SC}^{P, Q}_\eps$, the sample complexity of privately determining which distribution generates the samples. More precisely, we show in Lemma~\ref{lemma:BHT_SampleComplexity} that in the sequentially interactive (in fact, in the more general \textit{fully} interactive) setting
    $\mathsf{SC}^{P, Q}_\eps \gtrsim  \frac{e^{\eps}}{(e^\eps-1)^2} \max\big\{\frac{1}{\tv^2(P, Q)}, \frac{e^\eps}{H^2(P, Q)}\big\}$ 
    and  
    $\mathsf{SC}^{P, Q}_\eps \lesssim  \frac{e^{2\eps}}{(e^\eps-1)^2} \frac{1}{\tv^2(P, Q)}$ for any $\eps\geq 0$, where $H^2(P, Q)$ is the squared Hellinger distance between $P$ and $Q$. These bounds subsume and generalize the best existing result in \cite{Duchi_LDP_MinimaxRates} which indicates $\mathsf{SC}^{P, Q}_\eps = \Theta\big(\frac{1}{\eps^2\tv^2(P, Q)}\big)$ for sufficiently small $\eps$. Furthermore, they have recently been shown in \cite[Theorem 1.6]{Assadi_BHT} to be optimal (up to a constant factor) for any $\eps\ge 0$ if $P$ and $Q$ are binary. This, in fact, implies that (sequential or full) interaction does not help in the locally private hypothesis testing problem if $P$ and $Q$ are binary or if $\eps\leq 1$. Therefore, our results extend \cite[Theorem 5.3]{Joseph2019TheRO} that indicates the optimal mechanism is non-interactive for $\eps\leq 1$.
\end{enumerate}

\begin{table*}[htb]
\centering
\small
      \begin{tabular}{| c | c | c |c|}
      \hline
      {\textbf {Problem}} & {\textbf {UB}} & {\textbf {Previous LB}} & {\textbf {LB}} \\ \hline
      
       {\parbox[c][1.5cm]{4cm}{ \centering  \textbf {Entropy estimation}}} & 
       \parbox[c][1.3cm]{1.7cm}{ \centering N.A.} &
       \parbox[c][1cm]{1cm}{ \centering N.A.} 
       & \parbox[c][1cm]{5.2cm}{\centering$\min\big\{1, \frac{1}{n}\big[\frac{e^{\eps}+1}{e^{\eps}-1}\big]^2\big\}\log^2k$ \\\vspace{0.2cm}(Corollary~\ref{cor:entropt_estimation})} \\ \hline
       {\parbox[c][1.6cm]{4cm}{\centering  \textbf {Distribution estimation, \\ \centering  $\ell_h$-norm}}} &\parbox[c][1.6cm]{2.5cm}{\centering ${\frac{e^{\eps(1-1/h)} \Paren{e^\eps+d}^{1/h}} { \sqrt{n} (e^\eps-1)} }$ \\\vspace{0.2cm}\hspace{-0.15cm}\centering{(Theorem~\ref{thm:hadamard}) }}
       &\multicolumn{2}{c}{\parbox[c][1.5cm]{5.2cm}{\centering$ \min\Big\{1,  \frac{e^{\eps/2}  d^{1/h}} {\sqrt{n} (e^\eps-1) }, \Big[\frac{e^{\eps/2}}{\sqrt{n} (e^\eps-1)}\Big]^{1-1/h}\Big\}$ \\\vspace{0.2cm}\hspace{5pt}(Corollary~\ref{cor_probability_estimation}),~\cite{Jayadev_Unified} }}\vline \\\hline
       {\parbox[c][1.5cm]{4cm}{ \centering \textbf {Density estimation, \\ \centering $\ell_h$-norm, $\beta$-H\"older}}} & \parbox[c][1.3cm]{1cm}{\centering N.A. }& \parbox[c][1.3cm]{2.9cm}{\centering $(n\eps^2)^{\frac{-h\beta}{2\beta + 2}}$ for $\eps\leq 1$ \\ \vspace{0.2cm}\cite{LDP_density_estimation}}
       &\parbox[c][1.1cm]{3cm}{\centering $(ne^{-\eps}(e^\eps-1)^2)^{\frac{-h\beta}{2\beta + 2}}$ \\\vspace{0.2cm}\centering(Corollary~\ref{cor:DensityEstimation}) } \\\hline
       {\parbox[c][1.5cm]{4cm}{\centering \textbf {Gaussian location model, \\ \centering arbitrary loss}}} & \parbox[c][1.3cm]{1cm}{\centering N.A. }& \parbox[c][1cm]{1cm}{\centering N.A}
       &\parbox[c][1cm]{5.2cm}{\centering $\frac{\sqrt{d}}{e^2(V_d\Gamma(1+d))^{1/d}}\min\big\{1, \sqrt{\frac{\sigma^2 d}{n}}(\frac{e^\eps+1}{e^\eps-1})\big\}$ \\\vspace{0.2cm}(Corollary~\ref{cor:GaussianLocation_general}) } \\\hline
       {\parbox[c][1.5cm]{4cm}{\centering \textbf {Sample complexity of \\ \centering hypothesis testing}}} & \parbox[c][1.3cm]{2.1cm}{\centering $\frac{e^{2\eps}}{(e^\eps-1)^2} \frac{1}{\tv^2(P, Q)}$ \\\vspace{0.2cm}(Lemma~\ref{lemma:BHT_SampleComplexity})} & \parbox[c][1.3cm]{3cm}{\centering $\frac{1}{\eps^2 \tv^2(P, Q)}$ for $\eps\leq 1$\\\vspace{0.2cm} \centering\cite{Cannone_HypothesisTesting}}
       &\parbox[c][0.9cm]{5.2cm}{\centering\small{$\frac{e^\eps}{(e^\eps-1)^2}\max\big\{\frac{1}{\tv^2(P, Q)}, \frac{e^\eps}{H^2(P,Q)}\big\}$}\\\vspace{0.2cm} \centering (Lemma~\ref{lemma:BHT_SampleComplexity}) }\\\hline
       
      \end{tabular}
    \caption{\label{tab:pure} Summary of the minimax risks for $\eps$-LDP estimation, where we have omitted constants for all the results. For the distribution estimation with $\ell_h$-norm, our upper bound, built on Hadamard response mechanism discussed in Appendix~\ref{appen:hadamard}, is order optimal in $n$ and $d$ for the dense case unless $\eps \gtrsim \log d$. For the Gaussian location model, we consider the problem of privately estimating $\theta\in \Theta$ from $X^n\stackrel{\text{iid}}{\sim}\N(\theta, \sigma^2\mathnormal I_d)$. The result shown in this table assumes that $\Theta$ is the unit $\ell_2$-ball, where $V_d$ is the volume of the unit $\|\cdot\|$-ball (for arbitrary norm). Corollary~\ref{cor:GaussianLocation_general}, however, concerns with the general $\Theta$.  }
\end{table*}
\subsection{Additional Related Work}
Local privacy is arguably one of the oldest forms of privacy in statistics literature and dates back to Warner \cite{warner1965randomized}. This definition resurfaced in \cite{evfimievski2003limiting} and was adopted in the context of differential privacy as its local version. 
The study of statistical efficiency under LDP was initiated in \cite{Duchi_LDP_MinimaxRates, Duchi2016MinimaxOP} in the minimax setting and has since gained considerable attention.
While the original bounds on the private minimax risk in \cite{Duchi_LDP_MinimaxRates, Duchi2016MinimaxOP} were meaningful only in the high privacy regime (i.e., small $\eps$), the order optimal bounds were recently given for several estimation problems in \cite{Duchi_interactivity} for the general privacy regime. Interestingly, their technique relies on the decay rate of mutual information over a Markov chain, which is known to be equivalent to the contraction coefficient under KL-divergence \cite{Anantharam_SDPI}. However, their technique is quite different from ours in that it did not concern computing the contraction coefficient of an LDP mechanism. 

Among locally private statistical problems studied in the literature, two examples have received considerably more attention, namely, mean estimation and discrete distribution estimation.   For the first problem, Duchi et al.\ \cite{Duchi2016MinimaxOP} used \eqref{Duchi1_KL} to develop asymptotically optimal procedures for estimating the mean in the high privacy regime (i.e., $\eps<1$). For the high privacy regime (i.e., $\eps>1$), a new algorithm was proposed in \cite{Duchi_Federatedprotection} that is optimal and matches the lower bound derived in \cite{Duchi_interactivity} for interactive mechanisms. 
There has been more work on locally private mean estimation that studies the problem under additional constraints \cite{Ampli_Shuffling, cp_SGD,Shuffled_FL, Asoodeh_LDP_Equivalent, VQ_SGD,rohde2020, Jayadev_Unified, Jayadev_chiSquare_Contraction, LDP_Fisher, Optimal_mean_LDP, feldman2022private}. For the second problem, Duchi et al.\ \cite{Duchi_LDP_MinimaxRates} studied (non-interactive) locally private distribution estimation problem under  $\ell_1$ and $\ell_2$ loss functions and derived the first lower bound for the minimax risk, which was shown to be optimal~\cite{kairouz16_LDPEstimation} for high privacy regime. Follow-up works such as \cite{LDP_DistributionEstimation, LDP_Fisher, Jayadev_Unified,Optimal_compressionLDP_Feldman, Optimal_compressionLDP_Kairouz} characterized the optimal minimax rates for general $\eps$. Recently, \cite{Jayadev_Unified} derived a lower bound for $\ell_h$ loss with $h\geq 1$.    

The problem of locally private entropy estimation has received significantly less attention in the literature, despite the vast line of research on the non-private counterpart. The only related work in this area seems to be \cite{RenyiEntropy_LDP, Renyi-LDP_Lambda2} which studied estimating R\'enyi entropy of order $\lambda$ and derived optimal rates only when $\lambda>2$. Thus, the optimal private minimax rate seems to be still open. We remark that \cite{acharya2018inspectre} explicitly considered the problem of entropy estimation, but in the setting of central differential privacy.

\newhz{The closest work to ours are \cite{Asoodeh_LDP_Equivalent,Zamanlooy} which extensively studied  the contraction coefficient of LDP mechanisms under the hockey-stick divergence. More specifically, it was shown in \cite{Asoodeh_LDP_Equivalent} that $\sK$ is $\eps$-LDP \textit{if and only if} $\sE_{e^\eps}(P\sK\|Q\sK)$ the hockey-stick divergence between $P\sK$ and $Q\sK$ is equal to zero for any distributions $P$ and $Q$, and thus if and only if the contraction coefficient of $\sK$ under the hockey-stick divergence is zero.
By representing $\chi^2$-divergence in terms of the hockey-stick divergence, this result leads to a conceptually similar, albeit weaker, result as Theorem~\ref{Thm:Chi_TV}. }

\newhz{In~\cite{acharya2021differentially}, Acharya et al. introduced an information-theoretic toolbox to establish lower bounds for private estimation problems. However, they considered the threat model of central differential privacy, a totally different model from the local differential privacy considered in this work.}


\subsection{Notation}

We use upper-case letters (e.g., $X$) to denote random variables and  calligraphic letters to represent their support sets (e.g., $\X$). We write $X^n$ to denote $n$ random variables $X_1, \dots, X_n$. The set of all distributions on $\X$ is denoted by $\P(\X)$. 
A mechanism (or channel) $\sK:\X\to \P(\Z)$ is specified by a collection of distributions $\{\sK(\cdot|x)\in \P(\Z):x\in \X\}$. Given such mechanism $\sK$ and $P\in \P(\X)$, we denote by $P\sK$ the output distribution of $\sK$ when the input is distributed according to $P$, given by $P\sK(A)\coloneqq \int P(\text{d}x)\sK(A|x)$ for $A\subset \Z$. We use $\E_P[\cdot]$ to write the expectation with respect to $P$ and $[n]$ for an integer $n\geq 1$ to denote $\{1, \dots, n\}$.

\section{Preliminaries and Definitions}\label{sec:prelim}
In this section, we give basic definitions of $f$-divergence, contraction coefficients, and LDP mechanisms.
~\\
~\\
\textbf{$f$-Divergences and Contraction Coefficients.} 
Given a convex function $f:(0,\infty)\to\mathbb{R}$ such that $f(1)=0$, the $f$-divergence between two probability measures $P\ll Q$ is defined as \cite{Csiszar67, Ali1966AGC}
$D_f(P\|Q)\coloneqq \E_Q\big[f\big(\frac{\textnormal{d}P}{\textnormal{d}Q}\big)\big].$
Examples of $f$-divergences needed in the subsequent sections include: 
\begin{itemize}
    \item KL-divergence $D_\kl(P\|Q)\coloneqq D_f(P\|Q)$ for $f(t) = t\log t$,
    \item total-variation distance $\tv(P, Q)\coloneqq D_f(P\|Q)$ for $f(t)=\frac{1}{2}|t-1|$, 
    \item $\chi^2$-divergence $\chi^2(P\|Q)\coloneqq D_f(P\|Q)$ for $f(t) = t^2-1$,
    \item squared Hellinger distance $H^2(P,Q)\coloneqq D_f(P\|Q)$ for $f(t) = (1-\sqrt{t})^2$, and 
    \item hockey-stick divergence (aka $\sE_\gamma$-divergence \cite{Verdu:f_divergence}) $\sE_\gamma(P\|Q)\coloneqq D_f(P\|Q)$ for $f(t) = (t-\gamma)_+$ for some $\gamma\geq 1$, where $(a)_+\coloneqq \max\{a, 0\}$.
\end{itemize}
All $f$-divergences are known to satisfy the data-processing inequality. That is, for any channel $\sK:\X\mapsto \P(\Z)$, we have $D_f(P\sK\|Q\sK)\leq D_f(P\|Q)$ for any pair of distributions $(P, Q)$. However, this inequality is typically strict. One way to strengthen this inequality is to consider $\eta_f(\sK)$ the contraction coefficient of $\sK$ under $f$-divergence \cite{ahlswede1976} defined as  
\begin{equation}\label{eq:contractionCoefficient}
    \eta_f(\sK)\coloneqq \sup_{\substack{P, Q\in \P(\X):\\ D_f(P\|Q)\neq 0}}\frac{D_f(Q \sK\|P\sK)}{D_f(Q\|P)}.
\end{equation}
With this definition at hand, we can write $D_f(P\sK\|Q\sK)\leq \eta_f(\sK)D_f(P\|Q)$, which is typically referred to as the \textit{strong} data processing inequality. 
We will study in details contraction coefficients under KL-divergence, $\chi^2$-divergence, squared Hellinger distance, and total variation distance, denoted by $\eta_\kl(\sK)$, $\eta_{\chi^2}(\sK)$, $\eta_{H^2}(\sK)$, and $\eta_\tv(\sK)$, respectively, in the next section. We also need the following well-known fact about $\eta_\kl(\sK)$ \cite{Anantharam_SDPI}: 
\begin{equation}\label{eq:Contraction_MI}
    \eta_\kl(\sK) = \sup_{\substack{P_{XU}:\\ U-X-Z}}~\frac{I(U; Z)}{I(U; X)},
\end{equation}
where $\sK$ is the channel specifying $P_{Z|X}$, $I(A; B)\coloneqq D_\kl(P_{AB}\|P_AP_B)$ is the mutual information between two random variables $A$ and $B$, and $U-X-Z$ denotes the Markov chain in that order. Another important property of $\eta_\kl$ required in the proofs is its tensorization which is described in Appendix~\ref{Appen:tensotization}.  
~\\
~\\
\noindent \textbf{Local Differential Privacy}
%
A randomized mechanism $\sK:\X\to \P(Z)$ is said to be $\eps$-locally differentially private ($\eps$-LDP for short) for $\eps\geq 0$ if \cite{evfimievski2003limiting,Shiva_subsampling}  
$\sK(A|x)\leq e^\eps\sK(A|x'),$ for all $A\subset \Z$ and $x, x'\in \X$. 
Let $\Q_{\eps}$ be the collection of all $\eps$-LDP mechanisms $\sK$. It can be shown that LDP mechanisms can be equivalently defined in terms of the hockey-stick divergence: 
\begin{equation}\label{eq:LDP_HS}
    \sK\in \Q_\eps ~\Longleftrightarrow ~ \sEs(\sK(\cdot|x)\| \sK(\cdot|x'))=0, \forall x, x'\in \X. 
\end{equation}
\textcolor{black}{Arguably, the most known LDP mechanism is the binary randomized-response mechanism, introduced by Warner \cite{warner1965randomized}. For $\X = \{0, 1\}$, let the mechanism $\sK$ be defined as $\sK(\cdot|1) = \sBer(\kappa)$ and $\sK(\cdot|0) = \sBer(1-\kappa)$. It can be easily verified that this mechanism is $\eps$-LDP if $\kappa = \frac{e^\eps}{1+e^\eps}$.  In information theory parlance, the binary randomized response mechanism is a binary symmetric channel with crossover probability $\frac{1}{1+e^\eps}$.  A natural way to generalize this mechanism to the non-binary set is as follows. 
\begin{example}[($k$-ary randomized response)]\label{example_RR} Let $\X = \Z = [k]$. Let the mechanism $\sK$ be defined as
\begin{equation}\label{eq:k_array}
\sK(z|x) = \begin{cases}
    \frac{e^\eps}{e^\eps+ k-1}, & \text{if}~ z= x,\\
    \frac{1}{e^\eps+ k-1}, & \text{otherwise}.
\end{cases}
\end{equation}
It can be verified that $\sE_{e^\eps}(\sK(\cdot|x)\|\sK(\cdot|x')) = 0$ for all $x, x'\in [n]$, implying this mechanism is $\eps$-LDP. 
\end{example}
}

Suppose there are $n$ users, each in possession of a sample $X_i$, $i\in [n]\coloneqq \{1, \dots, n\}$. User $i$ applies a mechanism $\sK_i$ to generate $Z_i$ a privatized version of $X_i$. The collection of such mechanisms is said to be \textit{non-interactive} if $\sK_i$ is entirely determined by $X_i$ and independent of $(X_j,Z_j)$ for $j\neq i$. 
If, on the other hand, interactions between users are permitted, then  $\sK_i$ need not depend only on $X_i$.   In particular, the \textit{sequentially interactive} \cite{Duchi_LDP_MinimaxRates} setting refers to the case when  the input of $\sK_i$  depends on both $X_i$ and the outputs $Z^{i-1}$ of the $(i-1)$ previous mechanisms.

\section{Main Technical Results}
In this section, we present our main technical results.  First, we establish a tight upper bound on $\eta_{\chi^2}(\sK)$ for any $\eps$-LDP mechanisms by deriving an upper bound for $\chi^2(P\sK\|Q\sK)$ in terms of $\chi^2(P\|Q)$ for any pair of distributions $(P, Q)$. Interestingly, this upper bound is shown to hold for a large family of $f$-divergences, including KL-divergence and squared Hellinger distance. A similar result is known for total variation distance \citep[Corollary 11]{kairouz2014extremal_JMLR}: for any $\sK\in \Q_\eps$  
\begin{equation}\label{TV_SDPI}
     \eta_\tv(\sK)\leq \frac{e^{\eps}-1}{e^{\eps}+1}.
 \end{equation} 
It is known that $\eta_f(\sK)\leq \eta_\tv(\sK)$ for any channel $\sK$ and any $f$-divergences (see, e.g., \cite{COHEN_Paper, Raginsky_SDPI}). Thus, it follows from \eqref{TV_SDPI} that $\eta_f(\sK)\leq \frac{e^{\eps}-1}{e^{\eps}+1}$ for any $\sK\in \Q_\eps$.
This upper bound holds for general $f$-divergences, thus it is necessarily loose. The following theorem shows that a significantly tighter bound can be obtained for specific $f$-divergences. 
\begin{theorem}\label{thm_chiSDPI}
If $\sK$ is an $\eps$-LDP mechanism, then we have for any $\eps\geq 0$
\begin{equation}\label{thm1_general}
\eta_\kl(\sK) = \eta_{\chi^2}(\sK) = \eta_{H^2}(\sK) \leq \left[\frac{e^{\eps}-1}{e^{\eps}+1}\right]^2\eqqcolon \Upsilon_\eps.
\end{equation}
\end{theorem}

The upper bound given in this theorem is in fact tight, that is, there exists an $\eps$-LDP mechanism $\sK$ and a pair of distributions $(P, Q)$ such that $\chi^2(P\sK\|Q\sK) = \Upsilon_\eps\chi^2(P\|Q)$. To verify this, let $\sK$ be the randomized response mechanism, and consider $Q= \sBer(0.5)$ and $P= \sBer(\alpha)$ for some $\alpha\in [0,1]$.
In this case, it can be easily verified that $\chi^2(P\sK\|Q\sK) = 4\Upsilon_\eps \alpha^2$ and $\chi^2(P\|Q) = 4\alpha^2$. 
\textcolor{black}{Therefore, binary randomized response mechanism satisfies the inequality in \eqref{thm1_general} with equality. In addition to the tightness of Theorem~\ref{thm_chiSDPI}, this implies that the binary randomized response mechanism has the largest contraction coefficient among all $\eps$-LDP mechanisms. Considering $k$-ary randomized response mechanisms, it is therefore expected that the contraction coefficients decrease as $k$ increases, as proved next. 
\begin{proposition}[Contraction of $k$-ary randomized response]\label{proposition_RRContraction}
    Let $\sK$ be the $k$-ary randomized response mechanism defined in \eqref{eq:k_array}. Then, we have  
    $$\eta_{\chi^2}(\sK) = \frac{(e^\eps-1)^2}{(e^\eps+1)(e^\eps+k-1)}.$$
\end{proposition}
\begin{proof}
Recall that 
$$\eta_{\chi^2}(\sK) =  \sup_{\substack{P, Q\in \P(\X):\\ P\neq Q}}\frac{\chi^2(Q \sK\|P\sK)}{\chi^2(Q\|P)}.$$
It was recently shown in \cite{Polyanskiy_SDPI_Binary} that the above optimization can be restricted to pairs $P$ and $Q$ supported on two points in $\X$ for all mechanisms $\sK:\X\to \P(\Z)$ with countable $\X$. Therefore, due to the symmetry of $k$-ary randomized response mechanism, we can, without loss of generality, consider only $P = (\alpha, 1-\alpha, 0, \dots, 0)$ and $Q = (\beta, 1-\beta, 0, \dots, 0)$ in $\P([k])$ for $\alpha, \beta\in (0,1)$. In this case, we have 
$$\chi^2(P\sK\|Q\sK) = \frac{(e^\eps-1)^2 (e^\eps+1)(\alpha-\beta)^2}{(e^\eps+k-1)(\beta e^\eps+\bar\beta)(\beta+\bar\beta e^\eps)},$$ 
and
$$\chi^2(P\|Q) = \frac{(\alpha-\beta)^2}{\beta\bar\beta},$$
where $\bar\beta \coloneqq 1-\beta$.
Thus, we can write
\begin{equation}\label{Eq:Proof_RR}
    \eta_{\chi^2}(\sK) = \frac{(e^\eps-1)^2 (e^\eps+1)}{(e^\eps+k-1)} \sup_{\beta\in (0,1)}\frac{\beta\bar\beta}{(\beta e^\eps+\bar\beta)(\beta+\bar\beta e^\eps)}.
    \end{equation}
    Define $$g_\eps(\beta) = \frac{\beta\bar\beta}{(\beta e^\eps+\bar\beta)(\beta+\bar\beta e^\eps)}.$$ It can be verified that 
$$g'_\eps(\beta) = \frac{(1-2\beta)e^\eps}{(\beta e^\eps+\bar\beta)^2(\beta+\bar\beta e^\eps)^2},$$
thus $\beta\mapsto g_\eps(\beta)$ is increasing on $(0, \frac{1}{2}]$ and decreasing on $[\frac{1}{2}, 1)$. In particular, $g_\eps(\beta)$ attains its maximum at $\beta = \frac{1}{2}$, that is, 
$$\sup_{\beta\in (0,1)}\frac{\beta\bar\beta}{(\beta e^\eps+\bar\beta)(\beta+\bar\beta e^\eps)} = \frac{1}{(1+e^\eps)^2}.$$ Plugging this in \eqref{Eq:Proof_RR}, the desired result follows. 
\end{proof}
According to this result, the contraction coefficient of $k$-ary randomized response mechanism decreases as $k$ increases, see Fig~\ref{fig:RR}. 
\begin{figure}
    \centering
    \includegraphics[scale=0.5]{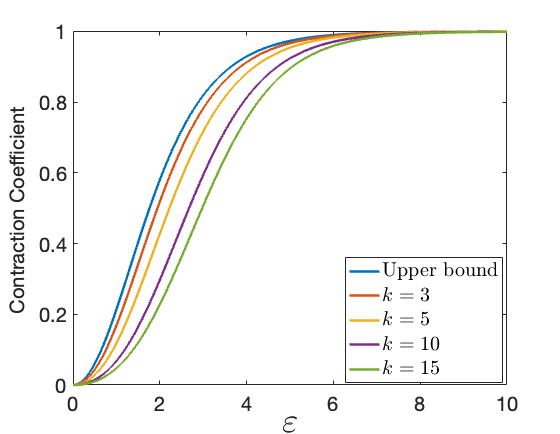}
    \caption{Contraction coefficient $\eta_{\chi^2}$ for $k$-ary randomized response mechanism with $k=3, 5, 10, 15$. Note that the contraction of binary randomized response coincides with the upper bound given in Theorem~\ref{thm_chiSDPI}. }
    \label{fig:RR}
\end{figure}
}

\begin{remark}
Proof of Theorem~\ref{thm_chiSDPI} reveals that the same result holds for a larger family of $f$-divergences. In fact, it can be shown that $\eta_f(\sK)\leq \Upsilon_\eps$ for $\sK\in \Q_\eps$ if $f$ is a non-linear ``operator-convex'' function, see e.g., \citep[Section III.C]{Raginsky_SDPI} and \citep[Theorem 1]{operator_Convex} for the definition of operator convex. The reason behind this generalization is that $\eta_f(\sK) = \eta_{\chi^2}(\sK)$ for all non-linear operator convex $f$, see e.g., \citep[Proposition 6]{Makur_SDPIjournal}, \citep[Proposition II.6.13 and Corollary II.6.16]{cohen1998comparisons}. 
\end{remark}



Theorem~\ref{thm_chiSDPI} turns out to be instrumental in studying several statistical problems under local privacy as discussed in Section~\ref{Sec:Application}. Nevertheless, it falls short in yielding a well-known fact about $\eps$-LDP mechanisms:  $\chi^2(P\sK\|Q\sK)<\infty$ even if $\chi^2(P\|Q) = \infty$. We address this issue in the next theorem which presents an upper bound for $\chi^2(P\sK\|Q\sK)$ in terms of  $\tv(P, Q)$, thus implying that $\chi^2(P\sK\|Q\sK)$ is always finite irrespective of $\chi^2(P\|Q)$. 
\begin{theorem}\label{Thm:Chi_TV}
If $\sK$ is an $\eps$-LDP mechanism, then
$$\chi^2(P\sK\|Q\sK)\leq \Psi_\eps\min\{4\tv^2(P, Q), \tv(P, Q)\},$$
for any pair of distributions $(P, Q)$ and $\eps\geq 0$, where 
\begin{equation}\label{eq:Psi_eps}
    \Psi_\eps\coloneqq e^{-\eps}(e^\eps-1)^2.
\end{equation}
\end{theorem}
The proof of this theorem relies partially on the proof of \citep[Proposition 8]{Duchi_FisherInfo}, which yields \eqref{Duchi3_tv}.
Nevertheless, Theorem~\ref{Thm:Chi_TV} is substantially stronger than \eqref{Duchi3_tv}, especially for $\eps\geq 1$. Notice that the upper bound in \eqref{Duchi3_tv} is of order $e^{\eps^2}$ for $\eps>1$ while Theorem~\ref{Thm:Chi_TV} gives a bound that scales as $e^\eps$. 
Note that since \newhz{$D_\kl(P\|Q)\leq \chi^2(P\|Q)$}, Theorem~\ref{Thm:Chi_TV} also gives an upper bound on $D(P\sK\|Q\sK)$ in terms of $\tv(P, Q)$ which is strictly stronger than \eqref{Duchi1_KL}.  
%

The upper bound in Theorem~\ref{Thm:Chi_TV} holds for all $\eps$-LDP mechanisms. However, for specific $\eps$-LDP mechanisms, one can achieve a slightly tighter upper bound. For instance, it can be shown that $\chi^2(P\sK\|Q\sK)\leq \Psi_\eps \tv^2(P, Q)$ for binary mechanisms (see Appendix~\ref{App:BinaryMechanism} for details).

\section{Applications}\label{Sec:Application}
In this section, we use the results presented in the previous section to examine several statistical problems under LDP constraint, including minimax estimation risks in Sections~\ref{sec:CRBound} to \ref{sec:MIM} and sample complexity of hypothesis testing in Section~\ref{sec:BHT}.  In all these applications, we allow our mechanisms to be sequentially interactive.

We first define private minimax estimation risk---the main quantity needed for most subsequent sections. 
Suppose $\{P_\theta\}_{\theta\in \Theta}$ for $\Theta\subseteq \R^d$ is a parametric family of probability measures on $\X$. If they are absolutely continuous, we denote their densities by $\{P_\theta\}_\theta$ as well. 
Let $X^n\coloneqq (X_1,\dots, X_n)$ be $n$ i.i.d.\ samples from $P_{\theta}$ that are distributed among $n$ users. User $i$ chooses $\sK_i\in \Q_\eps$ to generate $Z_i$ in a sequentially interactive manner, i.e., the distribution of $Z_i$ depends on $Z^{i-1}\coloneqq (Z_1, \dots, Z_{i-1})$. More specifically, $\sK_i$ receives $X_i$ and $Z^{i-1}$, and  generates $Z_i$. Thus, $Z_i\sim P_\theta\sK_i$ given a realization of $Z^{i-1}=z^{i-1}$. The goal is to estimate a function of $\theta$, denoted by $T(\theta)$, given the observation $Z^n$ via an estimator $\psi$. Invoking the minimax estimation framework to formulate this goal, we define private minimax estimation risk as
\begin{equation*}
    R^*(n, \Theta, \ell,\eps)\coloneqq \inf_{\sK_1,\dots, \sK_n\in \Q_\eps}\inf_{\psi} \sup_{\theta\in \Theta}\E\left[\ell(\psi(Z^n), T(\theta))\right],
\end{equation*}
where $\ell:\Theta\times \Theta\to \R^+$ is a loss function assessing the quality of an estimator. Note that $R^*(n, \Theta, \ell,\infty)$ corresponds to the non-private minimax risk. In the following sections, if  $T$ is not explicitly specified, then it is assumed to be identity, i.e., $T(\theta) = \theta$.
\subsection{Locally Private Fisher Information}\label{sec:CRBound}
Let the loss function be quadratic, i.e., $\ell = \ell_2$, and $I_{X}(\theta)$ be the Fisher information matrix  of $\theta$ given $X$ defined as  
\begin{equation}
    \newhz{I_{X}(\theta)\coloneqq \E[(\nabla\log P_\theta(X))(\nabla\log P_\theta(X))^{\mathsf{T}}]},
\end{equation}
where the gradient is taken with respect to $\theta$. It is well-known that an upper bound on the trace of the Fisher information matrix amounts to a lower bound on the minimax estimation risk associated with quadratic loss. This typically follows from Cram\'er-Rao bound (for unbiased estimators) or its Bayesian version known as van Trees inequality. 
Thus, it is desirable to obtain a sharp upper bound on $\mathsf{Tr}(I_{Z^n}(\theta))$. 
 
This has recently been noted in \cite{LDP_Fisher}, wherein several upper bounds for $\mathsf{Tr}(I_{Z^n}(\theta))$ were derived. However, those bounds only hold when $P_\theta$ satisfy some regularity conditions, namely $\E[(u^{\mathsf{T}}\nabla \log f_\theta(X))^2]$ is bounded for any unit vector $u\in \R^d$ or $\nabla\log f_\theta(X)$ is sub-Gaussian, where $f_\theta$ is the density of $P_\theta$ with respect to the Lebesgue measure.  These conditions are restrictive as they may not hold for general distributions. The following lemma gives an upper bound on $I_{Z^n}(\theta)$ that holds for any general $P_\theta$.   
\begin{lemma}\label{Lemm_FI_chi}
Let $X^n\stackrel{\textit{iid}}{\sim}P_\theta$ and $Z^n$ be the output of sequentially interactive mechanisms $\sK_1, \dots, \sK_n$ with $\sK_i\in \Q_\eps$ for $i\in [n]$. Then, we have  for every $\eps\geq 0$
$$I_{Z^n}(\theta)\preccurlyeq n\Upsilon_\eps I_X(\theta).$$
\end{lemma}
This lemma can be proved directly from Theorem~\ref{thm_chiSDPI} as follows. Let $\theta' = \theta + \zeta u$ for a unit vector $u\in \R^d$ and $\zeta\in \R$.  If $P_{\theta}$ and $P_{\theta'}$ are sufficiently close (i.e., $\zeta\to 0$), then it can be verified that for $n=1$
\begin{equation}\label{chi^2_1}
    \chi^2(P_\theta\sK\|P_{\theta'}\sK) = \zeta^2 u^{\mathsf{T}} I_Z(\theta) u + o(\zeta^2),
\end{equation}
and 
\begin{equation}\label{chi^2_2}
    \chi^2(P_\theta\|P_{\theta'}) = \zeta^2 u^{\mathsf{T}} I_X(\theta) u + o(\zeta^2).
\end{equation}
These identities, together with Theorem~\ref{thm_chiSDPI}, imply the desired upper bound on $I_{Z}(\theta)$. The proof for $n>1$ relies on the \textit{tensorization} property of the contraction coefficient discussed in Appendix~\ref{Appen:tensotization}. Next, we present a locally private version of the van Trees inequality. 
\begin{corollary}[Private van Trees Inequality]\label{cor:PrivatevanTrees}
For any $\eps\geq 0$ and $\Theta=[-B, B]^d$, we have 
\begin{equation*}
   R^*(n,\Theta, \ell^2_2,\eps)\geq 
    \frac{d^2}{n\Upsilon_\eps\sup_{\theta\in \Theta}\mathsf{Tr}(I_{X}(\theta)) + \frac{d\pi^2}{B^2}}.
\end{equation*}
\end{corollary}
The proof of this corollary is given in Appendix~\ref{app:vanTrees}, together with a locally private version of the Cram\'er-Rao bound. 
%
%
\subsection{Private Le Cam's Method: An Improved Version}\label{sec:LeCam}
In this section, we propose a private version of the Le Cam's method \cite{lecam1973}, which improves the existing one in the literature proved by Duchi et al. \cite{Duchi_LDP_MinimaxRates, Duchi2016MinimaxOP}. Their result states that for two families of distributions $P_{\Theta_1}=\{P_\theta, \theta \in \Theta_1\}$ and $P_{\Theta_2}=\{P_\theta, \theta \in \Theta_2\}$, with $\Theta_1, \Theta_2 \subseteq \Theta$ such that $\min_{\theta_1 \in \Theta_1,\theta_2 \in \Theta_2} \ell(T(\theta_1), T(\theta_2)) \ge \dist$, we have
\citep[Proposition 1]{Duchi2016MinimaxOP}
\begin{equation}\label{LeCam_Duchi}
   R^*(n, \Theta, \ell, \eps) \ge \frac{\dist}{2\sqrt{2}}\Big[\sqrt{2}- \sqrt{n}(e^\eps-1) \tv(P_1, P_2) \Big],  
\end{equation}
for any $P_1 \in P_{\Theta_1}$ and $P_2 \in P_{\Theta_2}$.
Applying Theorem~\ref{thm_chiSDPI} and Theorem~\ref{Thm:Chi_TV}, we can obtain a strictly tighter lower bound on $R^*(n, \Theta, \ell, \eps)$. 

\begin{theorem}[Improved private Le Cam's method]
\label{thm:private_le_cam}
Let $P_{\Theta_1}=\{P_\theta, \theta \in \Theta_1\}$ and $P_{\Theta_2}=\{P_\theta, \theta \in \Theta_2\}$, with $\Theta_1, \Theta_2 \subseteq \Theta$ such that $\min_{\theta_1 \in \Theta_1,\theta_2 \in \Theta_2} \ell(T(\theta_1), T(\theta_2)) \ge \dist$. Then, we have for any $P_1 \in P_{\Theta_1}$ and $P_2 \in P_{\Theta_2}$
\begin{align*}
    R^*(n, \Theta, \ell, \eps) &\ge \frac{\dist}{2\sqrt{2}} \Big[\sqrt{2}- \sqrt{n} \min \Big\{\sqrt{\Upsilon_\eps D_\kl(P_1\|P_2)}, 2\sqrt{\Psi_\eps} \tv(P_1, P_2),\sqrt{\Psi_\eps\tv(P_1, P_2)}\Big\}\Big].
\end{align*} 
\end{theorem}

Notice that since $\Psi_\eps < (e^\eps-1)^2$ for any $\eps>0$, this theorem yields a strictly better lower bound than \eqref{LeCam_Duchi}. In particular, it improves the dependency on $\eps$ from $e^{\eps}$ to $e^{\frac{\eps}{2}}$ for $\eps>1$. 

As an example of Theorem~\ref{thm:private_le_cam}, we next consider the locally private entropy estimation problem.  

\textbf{Entropy Estimation under LDP.} 
Consider the following setting: Given a parameter $\theta\in \Theta = [0, 1]^{k-1}$ satisfying $\sum_{i}\theta_i\leq 1$, we define the parametric distribution by $P_\theta = (\theta_1, \dots, \theta_{k-1}, \theta_k)$, where \newhz{$\theta_k = 1-\sum_i\theta_i$}. Thus, $P_\theta\in \P([k])$. We are interested in the entropy of $P_\theta$, i.e., \newhz{$H(P_\theta) = -\sum_{i=1}^k\theta_i\log\theta_i$.} \newhz{We design the following hypothesis testing problem:}  
let $P_1 = [\frac{2-\eta}{3}, \frac{1+\eta}{3(k-1)}, \dots, \frac{1+\eta}{3(k-1)}]$ for some $\eta\in [0, 2]$ and $P_2 = [\frac{2}{3}, \frac{1}{3(k-1)}, \dots, \frac{1}{3(k-1)}]$. It can be verified that $\left(H(P_2)-H(P_1)\right)^2 \ge \frac1{9} \eta^2 \log^2 (k-1)$ and $D_\kl(P_1\|P_2) \le \chi^2(P_1\| P_2) \le 2\eta^2$. Setting $\eta = \min\{1, \frac1{10\sqrt{n}}\frac{e^\eps+1}{e^\eps-1}\}$ and applying Theorem~\ref{thm:private_le_cam}, we arrive at the following lower bound. Our result improves the non-private lower bound by $1/\Upsilon_\eps$, which is at least a constant even when $\eps$ grows large. 

\begin{corollary}
\label{cor:entropt_estimation}
For the entropy estimation problem under LDP described above, we have for $k\geq 3$ and $\eps\geq 0$ 
$$R^*(n, [0, 1]^{k-1}, \ell_2,\eps)\ge \frac1{20} \min\Big\{1, \frac{1}{100 n\Upsilon_\eps}\Big\}\log^2(k-1).$$
\end{corollary}
It is worth pointing out that Butucea and Issarte \cite{RenyiEntropy_LDP} have recently studied estimating R\'enyi entropy of order $\lambda$ for any $\lambda \in (0, 1) \cup (1, \infty)$ under LDP
constraint. Specifically, they have established the minimax optimal rate $\Theta(\frac{1}{\eps^2 n})$ for $\lambda\geq  2$. However, they fell short of providing optimal rate for estimating entropy (i.e., the case where $\lambda \to 1$).
\subsection{Private Assouad's Method: An Improved Version}\label{sec:Assouad}
Although the Le Cam's method can provide sharp minimax rates for various problems, it is known to be constrained to applications that are reduced to binary hypothesis testing.
In this section, we provide a private version of the Assouad's method that is stronger than the existing one in \cite{Duchi_LDP_MinimaxRates, Duchi2016MinimaxOP}.  Let $\{P_\theta\}_{\theta \in \Theta}$ be a set of distributions indexed by $\mathcal{E}_k = \{\pm 1\}^k$ satisfying
\begin{equation}\label{Assoud_Cond}
 \ell\left(T(\theta_u), T(\theta_v) \right) \ge 2\tau \sum_{j=1}^k \mathbb{I}(u_j \neq v_j), ~~ \forall u,v  \in \mathcal{E}_k.   
\end{equation}
For each coordinate $i \in [k]$, we define the mixture of distributions obtained by averaging over distributions with a fixed value for the $j$-th position, i.e., 
\[
    P^n_{+j} \coloneqq \frac{1}{2^{k-1}}\sum_{v:v_j = +1} P^n_{\theta_v} 
    ~~~\text{and}~~~~P^n_{-j} \coloneqq \frac{1}{2^{k-1}}\sum_{v:v_j = -1} P^n_{\theta_v},
\]
where $P^n_{\theta_v}$ is the product distribution corresponding to $P_{\theta}$ when $\theta = \theta_v$ for $v\in \mathcal E_{k}$.
The non-private Assouad's method \cite{Yu1997} yields
\[
R^*(n, \Theta, \ell, \infty) \ge \frac{\tau}{2} \sum_{j=1}^k \left(1 - \tv\left(P^n_{+j}, P^n_{-j}\right) \right). 
\]
By applying Pinsker's inequality and \eqref{Duchi1_KL}, Duchi et al.\  \cite{Duchi2016MinimaxOP} extended this result to obtain a lower bound on the private minimax risk. Similarly, we apply Pinsker's inequality and Theorem~\ref{Thm:Chi_TV} to derive another bound for the private minimax risk which has a  stronger dependence on $\eps$.

\begin{theorem}[Improved private Assouad's method]
\label{thm:private_assouad}
Let the loss function $\ell$ satisfy \eqref{Assoud_Cond}, and define $P_{+j} = \frac{1}{2^{k-1}}\sum_{v:v_j = +1} P_{\theta_v}$ and $P_{-j} = \frac{1}{2^{k-1}}\sum_{v:v_j = -1} P_{\theta_v}$. Then, we have 
\[
  \newhz{R^*(n, \Theta, \ell, \eps) \ge \frac1{2} k\tau \Bigg[1-  \bigg[ \frac{2n \Psi_\eps}{k} { \sum_{j=1}^k \tv^2(P_{+j},P_{-j})}
  \bigg]^{\frac12} \Bigg]}.
\]
\end{theorem}
We apply this theorem to characterize lower bounds on the private minimax risk in the following two problems.  

\textbf{Private Distribution Estimation.} 
    Let $\Theta = \Delta_d = \{\theta\in [0, 1]^d:~\sum_{j=1}^d\theta_j = 1\}$ and each $X_i$ is distributed according to the multinomial distribution with parameter $\theta$ on $\X = [d]$. We assume that the loss function is the $\ell_h$-norm for some $h\geq 1$, i.e., $\ell(\theta, \hat\theta) = \|\theta-\hat\theta\|_h$. The private minimax risk for this problem has been extensively studied for $h=1$ and $h=2$, see e.g., \cite{Duchi_LDP_MinimaxRates,kairouz16_LDPEstimation,LDP_DistributionEstimation,   Disribution_estimation_hadamard,LDP_Fisher,Jayadev_unified_LDPConstraints,Acharya_discrete_estimationLDP}. The following corollary,  built on Theorem~\ref{thm:private_assouad}, gives a lower bound on the private minimax risk for all $h\geq 1$.
    \begin{corollary}\label{cor_probability_estimation}
    For any $h\geq 1$ and $\eps\geq 0$, we have 
\begin{align*}
    R^*(n,\Delta_d, \|\cdot\|_h, \eps)&\ge  \min\bigg\{1, {\frac{\sqrt{2}h}{h+1}} \Big[\frac1{2h+2}\Big]^{\frac1h} \frac{d^{1/h}}{\sqrt{n\Psi_\eps}},{\frac{\sqrt{2}h}{h+1}} \Big[\frac1{\sqrt{2}h}\Big]^{\frac1h} \Big[\frac{1}{\sqrt{n \Psi_\eps}}\Big]^{1-1/h}\bigg\}.
\end{align*}    
\end{corollary}

 This lower bound  matches (up to constant factors) with the upper bounds in~\cite{Disribution_estimation_hadamard,LDP_DistributionEstimation, Duchi_LDP_MinimaxRates,Jayadev_unified_LDPConstraints, bassily2019linear} for both $h=1$ and $h=2$, and thus is order optimal in these cases. Furthermore, compared to the best existing lower bound \cite{LDP_DistributionEstimation}, it improves the constants and applies to both non-interactive and sequentially interactive cases. 
 We remark that a lower bound was recently derived by Acharya et al. \citep[Theorem 5]{Jayadev_Unified} for general $h\geq 1$ which establishes the same order result as Corollary~\ref{cor_probability_estimation}. While both results have the same order analysis, our approach is more amenable to deriving constants. 

To further assess the quality of the lower bound in Corollary~\ref{cor_probability_estimation}, we obtain an upper bound on $R^*(n,\Delta_d, \|\cdot\|_h, \eps)$ by generalizing the Hadamard response~\cite{Disribution_estimation_hadamard} to $\ell_h$-norm with $h\ge2$ in Appendix~\ref{appen:hadamard}. Under some mild conditions, the upper bound coincides with the second term in Corollary~\ref{cor_probability_estimation}, with respect to the dependency on $d$ and $n$.

\textbf{Private Non-Parametric Density Estimation.} 
Suppose $X^n$ is a sequence of i.i.d.\ samples from a probability distribution on $[0,1]$ that has density $f$ with respect to the Lebesgue measure. Assume that $f$ is H\"older continuous with smoothness parameter $\beta\in  (0, 1]$ and constant $L$, i.e.
$$|f(x)-f(y)|\leq L |x-y|^\beta, ~~~~~~~~\forall x, y\in [0,1].$$
Let $\H^\beta_L([0,1])$ be the set of all such densities. We are interested in characterizing the private minimax rate in the sequentially interactive setting denoted by 
$$R^*(n,\H^\beta_L([0,1]),\|\cdot\|_h^h, \eps)\coloneqq \inf_{\sK_i\in \Q_\eps}\inf_{\hat f}\sup_{f}\E\big[\|f - \hat f \|_h^h\big],$$
where the expectation is taken with respect to the density $f\in \H^\beta_L([0,1])$ and also the mechanisms $\sK_1, \dots, \sK_n \in \Q_\eps$. 
The non-private minimax rate for this problem for $h=2$ is known to be $\Theta(n^{-\frac{2\beta}{2\beta+1}})$, see e.g., \citep[Theorem 4]{Fisher_LowerBound} for a more recent proof.  Butucea et al.~\cite{LDP_density_estimation} established a lower bound
on $R^*(n, \H^\beta_L([0,1]), \|\cdot\|_h^h, \eps)$ in the high privacy regime. In particular, it was shown  \citep[Proposition 2.1]{LDP_density_estimation} that 
\begin{equation}\label{Butucea1}
    R^*(n, \H^\beta_L([0,1]), \|\cdot\|_h^h, \eps)\gtrsim (n\eps^2)^{-\frac{h\beta}{2\beta+2}}.
\end{equation}
The proof of this result relies on \eqref{Duchi1_KL}, and thus it holds only for $\eps\leq 1$.  
Compared to the non-private minimax rate under $\ell_2$, this result indicates that the effect of local privacy for small $\eps$ concerns both the reduction of the effective sample size from $n$ to $n\eps^2$ and also change of the exponent of the convergence rate from $\frac{-2\beta}{2\beta + 1}$ to $\frac{-2\beta}{2\beta + 2}$. In the following corollary, we show that the same observation holds for all privacy regime by extending \eqref{Butucea1} to all $\eps\geq 0$.  More precisely, the privacy constraint causes the effective sample size to reduce from $n$ to $n\Psi_\eps$ and also the convergence rate to reduce to $\frac{-2\beta}{2\beta + 2}$ as before. 
\begin{corollary}\label{cor:DensityEstimation}
We have for $h\geq 1$ and $\eps\geq 0$
$$R^*(n, \H^\beta_L([0,1]), \|\cdot\|_h^h, \eps)\gtrsim (n\Psi_\eps)^{-\frac{h\beta}{2\beta+2}}.$$
\end{corollary}
This corollary is proved by incorporating Theorem~\ref{thm:private_assouad} into the classical framework that reduces the density estimation to a parameter estimation over a hypercube of a suitable dimension. 
Note that $\Psi_\eps \approx \eps^2$ for $\eps\leq 1$, thus Corollary~\ref{cor:DensityEstimation} recovers Butucea et al.'s result shown in \eqref{Butucea1}. 

\subsection{Locally Private Mutual Information Method}\label{sec:MIM}
Mutual information method has recently been proposed in \citep[Section 12]{Yihong_Lecturenote_Stat} as a systemic tool for obtaining lower bounds for non-private minimax risks with better constants than what would be obtained by Le Cam's and Assouad's methods. 
Let, for simplicity, $T$ be the identity function, i.e., $T(\theta) = \theta$. Moreover, suppose $\theta$ is distributed according to a prior $\pi\in \P(\Theta)$ and the loss function is the $r$th power of an \textit{arbitrary} norm over $\R^d$.  
Define the \textit{Bayesian} private risk as 
\begin{equation*}
    R^*_\pi(n, \Theta, \|\cdot\|^r, \eps) \coloneqq \inf_{\sK_1,\dots, \sK_n\in \Q_\eps}\inf_{\psi} ~\E_\pi\left[\|\psi(Z^n)- \theta\|^r\right]. 
\end{equation*} 
Notice that $R^*(n, \Theta, \|\cdot\|^r, \eps)\geq R^*_\pi(n, \Theta, \|\cdot\|^r, \eps)$ for any prior $\pi$. In the sequel, we expound an approach to lower bound  $R^*_\pi(n, \Theta, \|\cdot\|^r, \eps)$, which in turn yields a lower bound on $R^*(n, \Theta, \|\cdot\|^r, \eps)$.

 Fix $n$ mechanisms $\sK_1, \dots, \sK_n$ in $\Q_\eps$ that sequentially generate $Z^n$ and let $\hat \theta = \psi(Z^n)$ be an estimate of $\theta$ with the corresponding risk $\E_\pi[\|\theta-\hat\theta\|^r] \leq D$ for some $D\geq 0$. (We shall replace $D$ with $R^*_\pi(n, \Theta, \|\cdot\|^r, \eps)$ later.) We can clearly write  
\begin{equation*}
    I(\theta; \hat\theta) \geq  \inf_{P_{\hat\theta|\theta}}\{I(\theta; \hat\theta):\E_\pi[\|\theta-\hat\theta\|^r]\leq D\} \eqqcolon \mathsf{RDF}(\pi, D).
\end{equation*} 
Notice that the lower-bound is the definition of the rate-distortion function (RDF) evaluated at the distortion $D$, where the distortion measure is given by $\|\cdot\|^r$. On the other hand, the Markov chain $\theta-Z^n-\hat\theta$ and the data processing inequality imply $I(\theta; \hat\theta)\leq I(\theta; Z^n)$. Therefore, we have 
\begin{equation}\label{Bayesian1}
    \mathsf{RDF}(\pi, D) \leq I(\theta; Z^n). 
\end{equation}
Combining \eqref{eq:Contraction_MI} with the tensorization property of $\eta_\kl$, we can show that $$I(\theta; Z^n)\leq  I(\theta; X^n)\max_{i\in [n]} \eta_\kl(\sK_i),$$ see Appendix~\ref{appen:MIM} for details. Therefore, in light of Theorem~\ref{thm_chiSDPI} we have
\begin{equation}\label{Bayesian2}
    \mathsf{RDF}(\pi, D)\leq\Upsilon_\eps I(\theta; X^n). 
\end{equation}
If we could somehow analytically compute $\mathsf{RDF}(\pi, D)$ for a prior $\pi$, then \eqref{Bayesian2} would enable us to forge a relationship between $D$ and $I(\theta, X^n)$. This relationship is desirable as we can simply replace $D$ with $R^*_\pi(n, \Theta, \|\cdot\|^r, \eps)$. However, computing rate-distortion function is known to be notoriously difficult even for simple distortion measures. Nevertheless, we can invoke the Shannon Lower Bound (see, e.g., \cite{ShannonLowerBound} or \citep[Problem 10.6]{cover2012elements}) to find an asymptotically tight lower bound on $\mathsf{RDF}(\pi, D)$. This in turn leads to the following lower bound on $R^*_\pi(n, \Theta, \|\cdot\|^r, \eps)$.
\begin{theorem}[Locally private mutual information method]\label{thm:MIM}
Let $\theta\sim \pi$ for some $\pi\in \P(\Theta)$ and $X^n\stackrel{\text{iid}}{\sim}P_\theta$. For an arbitrary norm $\|\cdot\|$, we have 
\begin{equation*}
    R^*_\pi(n, \Theta, \|\cdot\|^r, \eps) \geq \frac{d}{r\cdot e \big[V_d\Gamma(1+d/r)\big]^{\frac rd}}~e^{H(\theta) - \Upsilon_\eps I(\theta; X^n)},
\end{equation*}
where $V_d$ is the volume of the unit $\|\cdot\|$-ball, $\Gamma(\cdot)$ is the Gamma function, and $H(\theta)$ is the entropy of $\theta\sim \pi$.
\end{theorem}
To obtain the best lower bound for $R^*(n, \Theta, \|\cdot\|^r, \eps)$ from Theorem~\ref{thm:MIM}, we need to pick a prior $\pi$ that maximizes $I(\theta; X^n)$.  This prior need not necessarily be supported on entire $\Theta$. An example of such prior selection is given for the Gaussian location model described next. 

\textbf{Private Gaussian Location Model.}
Suppose $P_\theta = \N(\theta, \sigma^2 \mathnormal{I}_d)$ for some $\sigma>0$, where $\theta\in \Theta$. 
Characterizing the minimax risk for estimating $\theta$ under LDP has been extensively studied for particular choices of loss function and $\Theta$. For instance,  $\|\cdot\| = \|\cdot\|_2$ and $\Theta = \text{unit}~\ell_\infty$-ball were adopted in \cite{Duchi_LDP_MinimaxRates, Duchi_FisherInfo, LDP_Fisher, Duchi_interactivity}), $\|\cdot\| = \|\cdot\|_2$ and $\Theta = \text{unit}~ \ell_2$-ball in \cite{Duchi_Federatedprotection} and $\|\cdot\| = \|\cdot\|_h$ for some $h>1$ and $\Theta = \text{unit}~\ell_\infty$-ball in \cite{Jayadev_Unified}. Theorem~\ref{thm:MIM} enables us to construct lower bounds on $R^*_\pi(n, \Theta, \|\cdot\|^r, \eps)$ for \textit{arbitrary} loss and \textit{arbitrary} $ \Theta$. For any such arbitrary subset $\Theta$ of $\R^d$, we define $\mathsf{rad}(\Theta)\coloneqq \inf_{y\in \R^d}\sup_{x\in \Theta}\|x-y\|_2$. 
\begin{corollary}[Private Gaussian location model]\label{cor:GaussianLocation_general}
Let $P_\theta = \N(\theta, \sigma^2 \mathnormal{I}_d)$ with $\sigma>0$ and $\theta\in \Theta$. Moreover, let $\|\cdot\|$ be an arbitrary norm over $\R^d$ and $\Theta$ be an arbitrary subset of $\R^d$ with a non-empty interior. Then, we have 
\begin{align*}
    R^*(n, \Theta,  \|\cdot\|^r, \eps)
     & \geq \frac{d^{1-r/2}}{re^2[V_d\Gamma(1+d/r)]^{r/d}}\left[\frac{V(\Theta)}{V_2(\Theta)}\right]^{r/d} \min\left\{\mathsf{rad}(\Theta)^r, \Big[\frac{\sigma^2 d}{n\Upsilon_\eps}\Big]^{r/2}\right\},
\end{align*}
where $V(\Theta)$ is the volume of $\Theta$ and $V_2(\Theta)$ is the volume of $\ell_2$-ball of radius $\mathsf{rad}(\Theta)$.
\end{corollary}
Instantiating this corollary, we may recover or generalize some existing lower bounds for Gaussian location models. For instance, for $\|\cdot\| = \|\cdot\|_2$, $r=2$, and $\Theta=\text{unit}~\ell_2$-ball, we have $V_d^{1/d}\asymp 1/\sqrt{d}$, $V(\Theta) = V_2(\Theta)$, $\mathsf{rad}(\Theta) = 1$, and $(\Gamma(1+\frac{d}{2}))^{1/d}\asymp \sqrt{d}$. It then follows from Corollary~\ref{cor:GaussianLocation_general} that  $R^*(n, \Theta,\|\cdot\|_2^2, \eps)\gtrsim    \min\big\{1, \frac{\sigma^2 d}{n \Upsilon_\eps}\big\}$ which is optimal for $\eps\leq 1$, as it matches the upper bounds in  \cite{Duchi_Federatedprotection}.  Also, for $\|\cdot\| = \|\cdot\|_h$ with $h\geq 1$, $r=1$, and $\Theta=\text{unit}~\ell_\infty$-ball, we have $V_d^{1/d}\asymp d^{-1/h}$, $V(\theta) = 2^d$, $V_2(\Theta) \asymp 1$, $\mathsf{rad}(\Theta) = 2\sqrt{d}$ and $\Gamma(1+d)\asymp d^d$. It then follows from Corollary~\ref{cor:GaussianLocation_general} that $R^*(n, \Theta,\|\cdot\|_h, \eps)\gtrsim    \min \big\{1, \sqrt{\frac{\sigma^2d^{2/h}}{n \Upsilon_\eps}}\big\}$, which generalizes \citep[Theorem 4]{Jayadev_Unified} from $\eps\leq 1$ to all $\eps\geq 0$.

\subsection{Binary Hypothesis Testing under LDP}\label{sec:BHT}
Consider the following typical setting of binary hypothesis testing: Given $n$ i.i.d.\ samples $X^n$ and two distributions $P$ and $Q$, we seek to determine which distributions generated $X^n$. That is, we wish to test the null hypothesis $H_0 = {P}$ against the alternative hypothesis $H_1 ={Q}$. To address the privacy concern, we take sequentially interactive mechanisms $\sK_1, \dots, \sK_n$ that generate $Z^n$. 
The goal is now to perform the above test given $Z^n$. Let $\phi:\Z^n\to \{0, 1\}$ be a test that accepts the null hypothesis if it is equal to zero.  For any such test $\phi$, define $A_n(\phi) = \{z^n\in \Z^n: \phi(z^n) = 1\}$. There are two error probabilities associated with $\phi$, namely, $P(A_n(\phi))$ and $1-Q(A_n(\phi))$.  
We say that this test privately distinguishes $P$ from $Q$ with sample complexity $n^*(\phi)$ if both $P(A_n(\phi))$ and $1-Q(A_n(\phi))$ are smaller than $1/10$ for every $n\geq n^*(\phi)$.
We then define the sample complexity of privately distinguishing $P$ from $Q$ as $$\mathsf{SC}^{P, Q}_\eps \coloneqq \inf_{\sK_1, \dots, \sK_n\in \Q_\eps}~\inf_{\phi:\Z^n\to \{0,1\}}~ n^*(\phi).$$ 
The characterization of sample complexity of hypothesis testing is well-understood in the non-private setting: The number of samples needed to distinguish $P$ from $Q$ is $\Theta(1/H^2(P,Q))$\footnote{This statement is folklore, but see, e.g., \cite{folklore_BHT_sample} for a simple proof.}. Under local privacy,  it has been shown in \cite{Duchi_LDP_MinimaxRates} that $\mathsf{SC}^{P, Q}_\eps = \Theta(1/\eps^2\tv^2(P, Q))$ for sufficiently small $\eps$. In the following lemma, we extend this result to any $\eps\geq 0$. 
\begin{lemma}\label{lemma:BHT_SampleComplexity}
Given $\eps\geq 0$ and two distributions $P$ and $Q$, we have\footnote{It can be shown that Lemma~\ref{lemma:BHT_SampleComplexity} holds in the most general setting, i.e., in the \textit{fully} interactive setting (aka the so-called blackboard model); see \cite{Joseph2019TheRO,Duchi_interactivity} for the formal definition.} 
\begin{equation*}
   \frac{4}{35}\max\left\{\frac{1}{\Upsilon_\eps H^2(P, Q)}, \frac{1}{2\Psi_\eps\tv^2(P, Q)}\right\} \le  \mathsf{SC}^{P, Q}_\eps \le \frac{2\log (5)}{\Upsilon_\eps\tv^2(P, Q)}. 
\end{equation*}
\end{lemma}
Our lower bound reveals an interesting phase transition: the sample complexity of the binary hypothesis testing appears to be dependent on the Hellinger distance instead of the total variation distance as $e^\eps \ge \Omega \left( \frac{H^2(P, Q)}{\tv^2(P, Q)} \right)$. {Furthermore, when $e^\eps$ is large, our result has made a constant-factor ($1/\Upsilon_\eps$) improvement compared to the non-private lower bound. }
\textcolor{black}{Recently, \newhz{Pensia} et al. \cite{Assadi_BHT} formalized these observations and demonstrated that the lower bound in Lemma~\ref{lemma:BHT_SampleComplexity} is in fact optimal (up to a constant factor) for any $\eps \ge 0$ if $P$ and $Q$ are binary, that is, 
$$\mathsf{SC}^{P, Q}_\eps \asymp \begin{cases}
    \frac{1}{\eps^2\tv^2(P, Q)}, & \text{if}~\eps\in (0, 1),\\
    \frac{1}{e^\eps\tv^2(P, Q)}, & \text{if}~e^\eps\in \Big[e, \frac{H^2(P, Q)}{\tv^2(P, Q)}\Big],\\
    \frac{1}{H^2(P, Q)}, & \text{if}~e^\eps > \frac{H^2(P, Q)}{\tv^2(P, Q)}.
\end{cases}$$
}

\small{
\bibliographystyle{plain}
\bibliography{reference}
}
\newpage
\normalsize
\appendices

\section{Tensorization of Contraction Coefficient}\label{Appen:tensotization}
Recall the definition of the contraction coefficient of $\sK$ under $f$-divergence:
\begin{equation}
    \eta_f(\sK)\coloneqq \sup_{\substack{P, Q\in \P(\X):\\ Q\neq P}}\frac{D_f(Q \sK\|P\sK)}{D_f(Q\|P)},
\end{equation}
that quantifies the extent at which data processing inequality can be improved. In this definition, the supremum is taken over both distributions $P$ and $Q$. Fixing the input distribution of $\sK$ in the above definition, we define the \textit{distribution-dependent} contraction coefficient as
\begin{equation}
    \eta_f(P, \sK)\coloneqq \sup_{\substack{Q\in \P(\X):\\ Q\neq P}}\frac{D_f(Q \sK\|P\sK)}{D_f(Q\|P)}.
\end{equation}
Clearly $\eta_f(\sK) = \sup_{P\in \P(\X)} \eta_f(P, \sK)$ and thus $\eta_f(P, \sK) \leq \eta_f(\sK)$ for any distribution $P$. Consider now $n$ distributions $P_1, \dots, P_n$ 
and denote by $P_1\otimes\dots \otimes P_n$ their product distribution. Also, consider $n$ mechanisms $\sK_1, \dots, \sK_n$ and denote by $\sK_1\otimes \dots \otimes \sK_n$ the corresponding mechanism obtained by composing them independently, i.e., $\sK_1\otimes \dots \otimes \sK_n: \X^n\to\P(\Z^n)$ defined by 
$$(\sK_1\otimes \dots \otimes \sK_n)(z^n|x^n) = \prod_{i=1}^n \sK_i(z_i|x_i).$$
An important question in information theory and statistics is to characterize the distribution-dependent contraction coefficient for $\eta_f(P_1\otimes\dots \otimes P_n, \sK_1\otimes \dots \otimes \sK_n)$ in terms of $\eta_f(P_i, \sK_i)$. It turns out if $f$ satisfies some regularity conditions then the corresponding  distribution-dependent contraction coefficient \textit{tensorizes}, that is  
\begin{equation}
    \eta_f(P_1\otimes\dots \otimes P_n, \sK_1\otimes \dots \otimes \sK_n) = \max_{i\in [n]} ~\eta_f(P_i, \sK_i).
\end{equation}
This result was first proved by Witsenhausen \cite{Witsenhausen_MC} for $\chi^2$-divergence and then by \cite{ahlswede1976} for KL-divergence.  The most general case was recently proved in Theorem 3.9 in~\cite{Raginsky_SDPI}. Thus, we have 
\begin{equation}
    \eta_\kl(P_1\otimes\dots \otimes P_n, \sK_1\otimes \dots \otimes \sK_n) = \max_{i\in [n]} ~\eta_\kl(P_i, \sK_i),
\end{equation}
and 
\begin{equation}
    \eta_{\chi^2}(P_1\otimes\dots \otimes P_n, \sK_1\otimes \dots \otimes \sK_n) = \max_{i\in [n]} ~\eta_{\chi^2}(P_i, \sK_i),
\end{equation}
In particular, we can write 
\begin{equation}\label{eq:tensorization_kl}
    \eta_\kl(P_1\otimes\dots \otimes P_n, \sK_1\otimes \dots \otimes \sK_n) \leq \max_{i\in [n]}~\eta_\kl(\sK_i),
\end{equation}
and 
\begin{equation}\label{eq:tensorization_chi}
    \eta_{\chi^2}(P_1\otimes\dots \otimes P_n, \sK_1\otimes \dots \otimes \sK_n) \leq \max_{i\in [n]} ~\eta_{\chi^2}(\sK_i).
\end{equation}

\section{Locally Private Cram\'er-Rao Bound and van Trees Inequality }\label{app:vanTrees}
\begin{proof}[Proof of Corollary~\ref{cor:PrivatevanTrees}]
Notice that the classical van Trees inequality is the Bayesian version of the Cram\'er-Rao bound. Let $\pi$ be the prior distribution on $\theta$ such that $\pi(\theta) = \prod_{j=1}^d\pi_j(\theta_j)$. Applying the multivariate version of the van Trees inequality proved in   \cite{vanTrees}, we obtain for any estimator $\psi$
\begin{equation}\label{CRBound33}
    \int\E[\|\psi(X^n)-\theta\|_2^2]\pi(\theta)\text{d}\theta\geq \frac{d^2}{\int \mathsf{Tr}(I_{X^n}(\theta))\pi(\theta)\text{d}\theta + \mathcal{J}(\pi)},
\end{equation}
where $\mathcal{J}(\pi)$ is the Fisher information associated with the prior $\pi$ defined as 
$$\mathcal{J}(\pi) = \sum_{j=1}^d\int\frac{(\pi'_j(\theta_j))^2}{\pi_j(\theta_j)}\text{d}\theta_j.$$
Since \eqref{CRBound33} is a lower bound on the minimax risk for any prior $\pi$, we pick the one that minimizes $\mathcal{J}(\pi)$. It is known that for $\Theta = [-B, B]^d$, the minimum $\mathcal{J}(\pi)$ is equal to $\frac{d\pi^2}{B^2}$, \citep[Sec. 2.7.3]{Tsybakov_Book} for details. Therefore, we arrive at the following non-private minimax risk 
\begin{equation}\label{CRBound11}
   R^*(n,[-B, B]^d, \ell^2_2,\infty)\geq \frac{d^2}{\sup_{\theta\in \Theta} \mathsf{Tr}(I_{Z^n}(\theta)) + \frac{d\pi^2}{B^2}}.
   \end{equation}
We remark that this inequality also appears in \citep[Section 2]{LDP_Fisher}.  To obtain a private version of the above lower bound, we can write 
\begin{equation}\label{CRBound1}
   R^*(n,[-B, B]^d, \ell^2_2,\eps)\geq \inf_{\sK_1,\dots, \sK_n\in \Q_\eps}
   \frac{d^2}{\sup_{\theta\in \Theta} \mathsf{Tr}(I_{Z^n}(\theta)) + \frac{d\pi^2}{B^2}}.
   \end{equation}
Applying Lemma~\ref{Lemm_FI_chi}, we conclude Corollary~\ref{cor:PrivatevanTrees}.
\end{proof}

One can similarly use Lemma~\ref{Lemm_FI_chi} to obtain a private version of Cram\'er-Rao bound. It follows from the multivariate version of the Cram\'er-Rao bound (see e.g., \citep[Theorem 11.10.1]{cover2012elements}) that for any unbiased estimator $\psi$
$$\sup_{\theta\in\Theta}\E[\|\psi(X^n)-T(\theta)\|_2] \geq  \sup_{\theta\in \Theta}(\nabla T(\theta))^{\mathsf{T}}I^{-1}_{X^n}(\theta)\nabla T(\theta).$$ 
Thus, if there exists any unbiased estimator, then 
$$R^*(n, \Theta, \ell_2,\infty)\geq \sup_{\theta\in \Theta}(\nabla T(\theta))^{\mathsf{T}}I^{-1}_{Z^n}(\theta)\nabla T(\theta),$$
and hence 
\begin{equation}\label{CRBound1}
   R^*(n, \Theta, \ell_2,\eps)\geq \inf_{\sK_1,\dots, \sK_n\in \Q_\eps}\sup_{\theta\in \Theta}(\nabla T(\theta))^{\mathsf{T}}I^{-1}_{Z^n}(\theta)\nabla T(\theta).
\end{equation}
Applying Lemma~\ref{Lemm_FI_chi}, we therefore conclude 
\begin{equation}\label{CRB_private}
   R^*(n, \Theta, \ell_2,\eps)\geq \frac{1}{n\Upsilon_\eps}\sup_{\theta\in \Theta}~(\nabla T(\theta))^{\mathsf{T}}I^{-1}_{X}(\theta)\nabla T(\theta).
\end{equation}
We must point out that this lower bound only holds if there exists an unbiased estimator $\psi$ for $T(\theta)$. However, it is not clear whether unbiased estimators always exist in local DP settings. Therefore, the applicability of \eqref{CRB_private} is limited. Nevertheless, we next apply this lower bound to the private entropy estimation problem, \textit{provided that there exists an unbiased entropy estimator.}  

Recall that we already studied the private entropy estimation problem in Section~\ref{sec:LeCam}, wherein we made use of Theorem~\ref{thm:private_le_cam} to derive a lower bound $R^*(n, \Theta, \ell_2,\eps)$. Here, we present an alternative proof. 

First notice that according to \eqref{CRB_private}, it suffices to compute the Fisher information matrix $I_X(\theta)$. follows
    \begin{equation}
        [I_X(\theta)]_{i,j} = -\E\left[\frac{\partial^2\log P_\theta(X)}{\partial \theta_i\partial \theta_j}\right] = \begin{cases}
        \frac{1}{\theta_i} +\frac{1}{\theta_k}, & \text{if}~ i = j,\\
        \frac{1}{\theta_k},& \text{if}~ i \neq j,
        \end{cases}
        \end{equation}
        and hence 
        \begin{equation}
            I_X(\theta) = \mathsf{diag}\left(\left[\frac{1}{\theta_1}, \dots, \frac{1}{\theta_{k-1}}\right]\right) + \frac{1}{\theta_k}\mathbf{1}_{k-1}\mathbf{1}_{k-1}^\mathsf{T},
        \end{equation}
        where $\mathsf{diag}([a_1, \dots, a_{k-1}])$ is a the diagonal matrix whose diagonal entries entries are given by $a_1, \dots, a_{k-1}$ and $\mathbf{1}_{k-1}$ is an all-one vector of size $k-1$. Invoking the Matrix Inversion Lemma, we obtain  
        \begin{equation*}
            I_X^{-1}(\theta) = \mathsf{diag}\left(\left[\theta_1, \dots, \theta_{k-1}\right]\right) + \left[\theta_1, \dots, \theta_{k-1}\right]^\mathsf{T}\left[\theta_1, \dots, \theta_{k-1}\right].
             \end{equation*}
             Next, we compute $\nabla T(\theta)$, where $T(\theta) = -\sum_{i=1}^k\theta_i\log\theta_i$ for each $\theta\in \Theta$. It can be easily verified that $\frac{\partial T}{\partial \theta_i} = \log\frac{\theta_k}{\theta_i}$ which, after straightforward manipulation, leads to
    \begin{equation}
        (\nabla T(\theta))^\mathsf{T}I_X^{-1}(\theta)\nabla T(\theta) = V(\theta),
    \end{equation}
    where $V(\theta)\coloneqq \var[\log P_\theta(X)]$ is the variance of $\log P_\theta(X)$ with $X\sim P_\theta$. In light of \eqref{CRB_private}, we  can therefore write   
    $$R^*(n, \Theta, \ell_2,\eps)\geq \left[\frac{e^{\eps}+1}{\sqrt{n}(e^{\eps}-1)}\right]^2 \sup_{\theta\in \Theta} V(\theta).$$
    We next show that $\sup_{\theta\in \Theta} V(\theta) = \Theta(\log^2 k)$. The upper bound comes from \citep[Eq. (464)]{Polyanskiy_ChannelCoding} that shows  $\sup_{\theta\in \Theta} V(\theta)\leq \log^2 k$. Conversely, it can be shown that $V(\theta) = \frac{2}{9}\log^2(2k-2)\geq \frac{2}{9}\log^2k$ for $\theta = \big(\frac{1}{3(k-1)}, \dots, \frac{1}{3(k-1)}\big)$. Therefore, we obtain 
    $$R^*(n, \Theta, \ell_2,\eps)\gtrsim  \min\Big\{1, \frac{1}{n\Upsilon_\eps}\Big\}\log^2k,$$
    which is the same as Corollary~\ref{cor:entropt_estimation} (up to constant factors).

\section{Missing Proofs}
\subsection{Proof of Theorem~\ref{thm_chiSDPI}}
Recall the definition of $\eta_f$ the contraction coefficient of $\sK$ under $f$-divergence in \eqref{eq:contractionCoefficient}, which is given in the following for convenience  
\begin{equation}\label{Eq:SDPI_f_Div}
    \eta_f(\sK)\coloneqq \sup_{\substack{P,Q\in \P(\X):\\ D_f(P\|Q)\neq 0}}\frac{D_f(P \sK\|Q\sK)}{D_f(P\|Q)}.
\end{equation}
We first prove $\eta_{\kl}(\sK)\leq \Upsilon_\eps$ for any $\sK\in \Q_\eps$. To this goal, we first need Theorem 1 in~\cite{Polyanskiy_SDPI_Binary} which states that the supremum in \eqref{Eq:SDPI_f_Div} is attained by binary distributions $P$ and $Q$. Moreover, it is known (Theorem 21 in \cite{Polyanskiy_SDPI_network}) that for any  binary-input channel $\sK$ (i.e., $\X=\{0,1\}$)\footnote{\newhz{It is worth mentioning that \cite[Theorem 21]{Polyanskiy_SDPI_network} was recently updated in \cite{Polyanskiy_SDPI_Binary} with the constant $\frac12$ being replaced with $\frac14$.}}, we have
\begin{equation}\label{UB_KL}
    \eta_\kl(\sK)\leq H^2(\sK(\cdot|0), \sK(\cdot|1))\left[1-\frac{1}{4}H^2(\sK(\cdot|0), \sK(\cdot|1))\right].
\end{equation}
Therefore, we can write for any general mechanism $\sK$
\begin{equation}\label{UB_KL2}
    \eta_\kl(\sK)\leq \sup_{x, x'\in \X}H^2(x, x')\left[1-\frac{1}{4}H^2(x, x')\right],
\end{equation}
where $H^2(x, x')\coloneqq H^2(\sK(\cdot|x), \sK(\cdot|x'))$ is the squared Hellinger distance between $\sK(\cdot|x)$ and $\sK(\cdot|x')$ for $x, x'\in \sK$. Since the squared Hellinger distance takes values in $[0,2]$ and the mapping $t\mapsto t(1-\frac{1}{4}t)$ in increasing on  $[0,2]$, an upper bound on $H^2(\sK(\cdot|x), \sK(\cdot|x'))$ for $x, x'\in \X$ leads to an upper bound on $\eta_\kl(\sK)$.  
To this goal, we invoke \eqref{eq:LDP_HS} to write 
\begin{equation}\label{H_HS}
    \sup_{\sK\in \Q_\eps}\sup_{x, x'\in \X}H^2(\sK(\cdot|x), \sK(\cdot|x'))\leq  \sup_{\substack{M, N\in\P(\Z)\\\sEs(M\|N)=0\\\sEs(N\|M)=0}}H^2(M, N).
\end{equation}
Note that from Equation (429) in~\cite{Verdu:f_divergence}, we have for any pair of distributions $(M,N)$
\begin{equation}\label{H_integral}
    H^2(M,N) =  \frac{1}{2}\int_{1}^\infty\left[\sE_\gamma(M\|N)+\sE_\gamma(N\|M)\right]\gamma^{-\frac{3}{2}}\text{d}\gamma. 
\end{equation}
If $M$ and $N$ satisfy $\sEs(N\|M)=0$ and $\sEs(M\|N)=0$,  then the monotonicity of $\gamma\mapsto \sE_\gamma(M\|N)$ implies that $\sE_\gamma(N\|M)=0$ and $\sE_\gamma(M\|N)=0$ for all $\gamma\geq e^\eps$. Consequently, we obtain from  \eqref{H_integral} that
\begin{equation}\label{H_integral2}
    H^2(M,N) =  \frac{1}{2}\int_{1}^{e^\eps}\left[\sE_\gamma(M\|N)+\sE_\gamma(N\|M)\right]\gamma^{-\frac{3}{2}}\text{d}\gamma. 
\end{equation}
The convexity of $\gamma\mapsto \sE_\gamma(M\|N)$ (see e.g., Proposition 4 in~\cite{E_gamma}) and the fact that $\sE_1(M\|N) = \tv(M,N)$ indicate that \begin{equation}
    \sE_\gamma(M\|N)\leq \left[\frac{e^\eps-\gamma}{e^\eps-1}\right]\tv(M,N),
\end{equation} 
for all $\gamma\leq e^\eps$. Plugging this into \eqref{H_integral2}, we obtain 
\begin{align}
    H^2(M,N) &\leq  \frac{\tv(M,N)}{e^\eps-1}\int_{1}^{e^\eps}\left[e^\eps-\gamma\right]\gamma^{-\frac{3}{2}}\text{d}\gamma\\
    & = 2\frac{(e^{\eps/2}-1)^2}{e^\eps-1} \tv(M,N)\label{H_max2}
\end{align}
Next, we derive an upper bound for $\tv(M, N)$ when $\sEs(N\|M)=0$ and $\sEs(M\|N)=0$:
\begin{equation}\label{TV_max}
   \sup_{\substack{M, N\\\sEs(M\|N)=0\\\sEs(N\|M)=0}} \tv(M, N).
\end{equation}
First, we show that this supremum is attained with binary distributions. To this goal, define $\phi:\X\to \{0,1\}$ as 
\begin{equation}
    \phi(x) = 
    \begin{cases}
    1, & \text{if}~\text{d}M(x)\geq \text{d}N(x),\\
    0, & \text{if}~\text{d}M(x)< \text{d}N(x).
    \end{cases}
    \end{equation}
Let also $M_\mathsf{b}$  and $N_\mathsf{b}$ be the Bernoulli distributions induced by  push-forward of $M$ and $N$ via $\phi$. It can be verified that $\tv(M, N) = \tv(M_\mathsf{b}, N_\mathsf{b})$. Moreover, due to the data-processing inequality, we have 
$\sEs(M_\mathsf{b}\|N_\mathsf{b}) = \sEs(N_\mathsf{b}\|M_\mathsf{b})=0$. Hence, we can write
\begin{align}
   \sup_{\substack{M, N\\\sEs(M\|N)=0\\\sEs(N\|M)=0}} \tv(M, N) &= \sup_{\substack{p, q\in [0, 1]\\\sEs(\mathsf{Ber}(p)\|\mathsf{Ber}(q))=0\\\sEs(\mathsf{Ber}(q)\|\mathsf{Ber}(p))=0}} \tv(\mathsf{Ber}(p), \mathsf{Ber}(q)) \nonumber\\
   & = \sup_{\substack{p, q\in [0, 1]\\ q \leq p\leq \min\{qe^\eps, qe^{-\eps}+1-e^{-\eps}\}}} (p-q)\nonumber\\
   & = e^{-\eps}\frac{(e^\eps-1)^2}{e^\eps-e^{-\eps}},\label{TV_max2}
\end{align}
where $\mathsf{Ber}(q)$ denotes the Bernoulli distribution for $q\in [0, 1]$ and the last equality comes from a basic linear programming problem.  

Plugging \eqref{TV_max2} into \eqref{H_max2}, we obtain 
\begin{equation}
    H^2(M, N)\leq 2\frac{(e^{\eps/2}-1)^2(1-e^{-\eps})}{e^\eps-e^{-\eps}},
\end{equation}
for any pair of distributions $M$ and $N$ satisfying $\sEs(M\|N)=\sEs(N\|M)=0$.
Therefore, according to \eqref{UB_KL2}, we have 
\begin{equation}\label{KLproof}
    \eta_\kl(\sK)\leq \frac{(e^{\eps}-1)^2}{(e^{\eps}+1)^2},
\end{equation}
for any $\sK\in \Q_\eps$, which is what we wanted to show. 

Next we prove that the similar result holds for $\eta_{\chi^2}(\sK)$ and $\eta_{H^2}(\sK)$. To do so, we note that (see e.g., Proposition II.6.13 and Corollary II.6.16 in~\cite{cohen1998comparisons}, Section III.C in~\cite{Raginsky_SDPI} and Theorem 1 in ~\cite{operator_Convex})
$\eta_f(\sK) = \eta_{\chi^2}(\sK)$ for all nonlinear and operator convex\footnote{The definition of operator convex is quite involved and we refer the readers to Section III.C in ~\cite{Raginsky_SDPI} for its definition.} $f$, e.g., for KL-divergence and for squared Hellinger distance. Therefore, we can write 
$$\eta_{\chi^2}(\sK) = \eta_{H^2}(\sK) = \eta_\kl(\sK),$$
for any mechanism $\sK$. This, together with \eqref{KLproof}, implies that $\eta_{\chi^2}(\sK) = \eta_{H^2}(\sK) = \Upsilon_\eps$ for all $\eps$-LDP mechanisms $\sK$. 

\subsection{Proof of Theorem~\ref{Thm:Chi_TV}}\label{appen:Chi_TV}
First, we notice that \cite[Proposition 8]{Duchi_FisherInfo} yields
\begin{equation}\label{eq:Duchi_chiUB}
     \chi^2(P\sK\|Q\sK)\leq 4 \sup_{x, x'\in \X} \chi^2(\sK(\cdot|x)\|\sK(\cdot|x')) \tv^2(P, Q). 
\end{equation}
Thus, an upper bound on $\chi^2(\sK(\cdot|x)\|\sK(\cdot|x'))$ leads to an upper bound on $\chi^2(P\sK\|Q\sK)$. We now derive an upper bound for $\sup_{x, x'\in \X} \chi^2(\sK(\cdot|x)\|\sK(\cdot|x'))$. 
To this goal, first note that, we can write analogously to \eqref{H_HS}
\begin{equation}
    \sup_{x, x'\in \X} \chi^2(\sK(\cdot|x)\|\sK(\cdot|x'))\leq \sup_{\substack{M, N\in \P(\Z)\\\sEs(M\|N)=0\\\sEs(N\|M)=0}}\chi^2(M\|N).
\end{equation}
To solve the latter optimization problem, we resort to the integral representation of $\chi^2$-divergence in term of $\sE_\gamma$-divergence (see e.g., Equation (430) in~\cite{Verdu:f_divergence})
\begin{equation}\label{integral_Chi2}
    \chi^2(M\|N) = 2\int_1^{\infty} \left[\sE_\gamma(M\|N) + \gamma^{-3}\sE_\gamma(N\|M)\right]\text{d}\gamma.
\end{equation}
Let $M$ and $N$ be distributions on $\Z$ satisfying $\sEs(M\|N)=0$ and $\sEs(N\|M)=0$. The monotonicity and convexity  of $\gamma\mapsto \sE_\gamma(M\|N)$ imply that $\sE_\gamma(M\|N)=\sE_\gamma(N\|M)=0$ for all $\gamma\geq e^\eps$ and $\sE_\gamma(M\|N)\leq \frac{\tv(M, N)(e^\eps-\gamma)}{e^\eps-1}$ for all $\gamma\leq e^\eps$. Thus, it follows from \eqref{integral_Chi2} 
\begin{align}
    \chi^2(M\|N) &\leq  \frac{2\tv(M, N)}{e^\eps-1}\int_1^{e^\eps} (e^\eps-\gamma)(1+ \gamma^{-3})\text{d}\gamma\label{integral_UB_chi}\\
    &=e^{-\eps}(e^\eps-1)(e^\eps+1) \tv(M, N)\nonumber\\
    & \leq e^{-\eps}(e^\eps-1)^2,\nonumber
\end{align}
where the last inequality follows from \eqref{TV_max2}. Plugging this upper bound into \eqref{eq:Duchi_chiUB}, we obtain 
\begin{equation}\label{eq:Duchi_chiUB2}
    \chi^2(M\|N)\leq 4  e^{-\eps}(e^\eps-1)^2 \tv^2(P, Q).
\end{equation}

We now prove the second part 
\begin{equation}
    \chi^2(P\sK\|Q\sK)\leq e^{-\eps}(e^\eps-1)^2\tv(P, Q).
\end{equation}
Note that, we can write from \eqref{integral_UB_chi} 
\begin{align}
    \chi^2(P\sK\|Q\sK) &\leq  \frac{2\tv(P\sK, Q\sK)}{e^\eps-1}\int_1^{e^\eps} (e^\eps-\gamma)(1+ \gamma^{-3})\text{d}\gamma\nonumber\\
    &=e^{-\eps}(e^\eps-1)(e^\eps+1) \tv(P\sK, Q\sK)\nonumber\\
    & \leq e^{-\eps}(e^\eps-1)^2\tv(P, Q),\label{thm2_proof_last2}
\end{align}
where the last inequality follows from \eqref{TV_SDPI}. Combining \eqref{eq:Duchi_chiUB2} and \eqref{thm2_proof_last2}, we drive the desired result.

\subsection{Binary Mechanism} \label{App:BinaryMechanism}
 Consider the binary mechanism $\sK:\X\to \P(\{0,1\})$ given by 
\begin{equation}\label{binary_mech}
    \sK(0|x) = 
\begin{cases}
\frac{e^\eps}{1+e^\eps}, & \text{if} ~P(x)\geq Q(x),\\
\frac{1}{1+e^\eps}, & \text{if} ~P(x)< Q(x).
\end{cases}
\end{equation}
The following proposition shows that the constant $4$ in Theorem~\ref{Thm:Chi_TV} can be replaced with $1$ for the binary mechanism. 
\begin{proposition}\label{Prop:Chi_TV}
For the binary mechanism, we have for any $\eps\geq 0$
$$\chi^2(P\sK\|Q\sK)\leq \Psi_\eps \tv^2(P, Q).$$
\end{proposition}
\begin{proof}
Note that for any $\alpha, \beta\in [0,1]$
$$\chi^2(\sBer(\alpha)\|\sBer(\beta)) = \frac{(\alpha-\beta)^2}{\beta\bar\beta},$$
where $\bar\beta\coloneqq 1-\beta$. 

Let $A=\{x\in \X: P(x)\geq Q(x)\}$. Since $\sK$ is a binary mechanism, it can be shown that $P\sK\sim \sBer(\zeta P(A^c)+\bar\zeta P(A))$ and similarly $Q\sK\sim \sBer(\zeta Q(A^c)+\bar\zeta Q(A))$, where $\zeta=\frac{e^\eps}{1+e^\eps}$ and $A^c$ is the complement of $A$.Thus, we have 
\begin{align*}
\chi^2(P\sK\|Q\sK) &= \frac{(P(A)-Q(A))^2 (2\zeta-1)^2}{(\zeta Q(A^c)+\bar\zeta Q(A))(\zeta Q(A)+\bar\zeta Q(A^c))}.
\end{align*}
Note that by definition $P(A)-Q(A) = \tv(P\|Q)$. Also, it can be easily shown that the denominator is greater than $\zeta\bar\zeta$. Thus, we can write
\begin{align*}
\chi^2(P\sK\|Q\sK) &\leq  \frac{(2\zeta-1)^2}{\zeta\bar\zeta}\tv^2(P, Q)\\
& = e^{-\eps}(e^\eps-1)^2\tv^2(P, Q).
\end{align*}
\end{proof}
\subsection{Proof of Lemma~\ref{Lemm_FI_chi}}
First, suppose $n=1$. Fix $\theta\in \Theta$ and $\theta' = \theta + \zeta u$ for a unit vector $u\in \R^d$ and $\zeta\in \R$. In light of Theorem~\ref{thm_chiSDPI}, we have for each $\sK\in \Q_\eps$
\begin{equation}
    \chi^2(P_\theta\sK\|P_{\theta'}\sK)\leq \left[\frac{e^{\eps}-1}{e^{\eps}+1}\right]^2 \chi^2(P_\theta\|P_{\theta'}).
\end{equation}
Plugging this inequality in \eqref{chi^2_1} and \eqref{chi^2_2} and letting $\zeta\to 0$, we obtain 
\begin{equation}
    I_Z(\theta)\preccurlyeq \left[\frac{e^{\eps}-1}{e^{\eps}+1}\right]^2  I_X(\theta),
\end{equation}
proving the desired result for $n=1$. For the general case $n>1$, 
we consider the tensorization property of the distribution-dependent contraction coefficient of $\chi^2$-divergence, described in Appendix~\ref{Appen:tensotization}. 
Let $P_\theta^{\otimes n}$ denote the distribution of $n$ i.i.d.\ samples from $P_\theta$ and $\sK^n = \sK_1, \dots, \sK_n$ denote the sequentially interactive mechanism obtained from $n$ mechanisms $\sK_1, \dots, \sK_n$. It follows from  \eqref{eq:tensorization_chi} that 
\begin{equation}
    \eta_{\chi^2}(P_\theta^{\otimes n},\sK^n) = \max_{i\in [n]} ~\eta_{\chi^2}(P_\theta,\sK_i).
\end{equation}
Also, similar to \eqref{chi^2_1} and \eqref{chi^2_2}, we can write
\begin{equation}\label{chi^2n_1}
    \chi^2(P_\theta^{\otimes n}\sK^n\|P_{\theta'}^{\otimes n}\sK^n) = \zeta^2 u^{\mathsf{T}} I_{Z^n}(\theta) u + o(\zeta^2),
\end{equation}
and 
\begin{equation}\label{chi^2n_2}
    \chi^2(P_\theta^{\otimes n}\|P_{\theta'}^{\otimes n}) = \zeta^2 u^{\mathsf{T}} I_{X^n}(\theta) u + o(\zeta^2).
\end{equation}
Thus, if each $\sK_i\in \Q_\eps$, then we can write 
\begin{align*}
    I_{Z^n}(\theta)&\preccurlyeq  I_{X^n}(\theta) \eta_{\chi^2}(P_\theta^{\otimes n},\sK^n) \\ 
    & = I_{X^n}(\theta) \max_{i\in [n]} ~\eta_{\chi^2}(P_\theta,\sK_i)\\
    &\preccurlyeq  I_{X^n}(\theta) \max_{i\in [n]} ~\eta_{\chi^2}(\sK_i)\\
    & \preccurlyeq  \left[\frac{e^{\eps}-1}{e^{\eps}+1}\right]^2 I_{X^n}(\theta), 
\end{align*}
where the third step follows from the fact that $\eta_{\chi^2}(P, \sK)\leq \eta(\sK)$ for any distribution $P$ and mechanism $\sK$, and the last step is due to Theorem~\ref{thm_chiSDPI}. The desired result then follows immediately by noticing  $I_{X^n}(\theta) = nI_X(\theta)$.

\subsection{Proof of Theorem~\ref{thm:private_le_cam}}   

According to the classical non-private Le Cam's method, for any families of distributions $P_{\Theta_1}=\{P_\theta, \theta \in \Theta_1\}$ and $P_{\Theta_2}=\{P_\theta, \theta \in \Theta_2\}$, with $\Theta_1, \Theta_2 \subseteq \Theta$ and any loss function $\ell$ satisfying 
$$\min_{\theta_1 \in \Theta_1,\theta_2 \in \Theta_2} \ell(T(\theta_1), T(\theta_2)) \ge \dist,$$ 
we have 
\[
  R^*(n, \Theta, \ell, \infty) \ge \frac{\dist}{2\sqrt{2}}\left(\sqrt{2}- \sqrt{D_\kl(P^{\otimes n}_1\|P^{\otimes n}_2)} \right),
\]
for any $P_1 \in P_{\Theta_1}$ and $P_2 \in P_{\Theta_2}$, where $P_1^{\otimes n}$ and $P_2^{\otimes n}$ denote the product distribution corresponding to $P_1$ and $P_2$, respectively.
It follows from this result that in the sequentially interactive setting, we have 
\begin{equation}\label{eq:LecamProof0}
    R^*(n, \Theta, \ell, \eps) \ge \inf_{\sK_1, \dots, \sK_n\in \Q_\eps} \frac{\dist}{2\sqrt{2}}\left(\sqrt{2}- \sqrt{D_\kl(P_1^{\otimes n}\sK^n\|P_2^{\otimes n}\sK^n)} \right),
\end{equation}
where $\sK^n=\sK_1, \dots, \sK_n$ denotes the sequentially interactive mechanism obtained from $n$ mechanisms $\sK_1, \dots, \sK_n$.
Note that according to the tensorization property of $\eta_\kl$ and Theorem~\ref{thm_chiSDPI}, we have 
\begin{equation}\label{eq:LecamProof1}
    D_\kl(P_1^{\otimes n}\sK^n\|P_2^{\otimes n}\sK^n)\leq \Upsilon_\eps D_\kl(P_1^{\otimes n}\|P_2^{\otimes n}) = n \Upsilon_\eps D_\kl(P_1\|P_2).
\end{equation}
On the other hand, applying chain rule of KL-divergence and Theorem~\ref{Thm:Chi_TV} (similar to Proposition 1 in~\cite{Duchi2016MinimaxOP}), we obtain 
\begin{align}
    D_\kl(P_1^{\otimes n}\sK^n\|P_2^{\otimes n}\sK^n) &\leq  \sum_{i=1}^n\int D_\kl(P_1\sK_i(\cdot|z^{i-1})\|P_2\sK_i(\cdot|z^{i-1}))\text{d}P(z^{i-1}) \nonumber\\
    & \leq  n \Psi_\eps \min\{4\tv^2(P_1, P_2), \tv(P_1, P_2)\},\label{eq:LecamProof2}
\end{align}
where in the first step $P$ denotes the distribution of $Z^{i-1}$.
Plugging \eqref{eq:LecamProof1} and \eqref{eq:LecamProof2} into \eqref{eq:LecamProof0}, we arrive at the desired result.

\subsection{Proof of Theorem~\ref{thm:private_assouad}}
By the classical Assouad's method, we can write 
\[
R^*(n, \Theta, \ell, \eps) \ge \frac{\tau}{2} \sum_{j=1}^k \left(1 - \tv\left(M^n_{+j}, M^n_{-j}\right) \right),
\]
where $M^n_{+j}$ and $M^n_{-j}$, which are the distributions of $P^n_{+j}$ and $P^n_{-j}$ after the channel.
By Pinsker's inequality and the Cauchy-schwartz inequality,
\begin{align}
\label{equ:assouad_proof}
\sum_{j=1}^k \tv\left(M^n_{+j}, M^n_{-j}\right) \le \sqrt{k} \cdot \sqrt{\sum_{j=1}^k \tv^2\left(M^n_{+j}, M^n_{-j}\right)} \le \sqrt{\frac{{k}}{2}} \cdot \sqrt{\sum_{j=1}^k D_\kl\left(M^n_{+j}, M^n_{-j}\right)}.
\end{align}

Next we upper bound $D_\kl(M^n_{+j}\|M^n_{-j})$ for each $j\in [k]$. To this end, note that 
\begin{equation}
    D_\kl(M^n_{+j}\|M^n_{-j}) = \sum_{i=1}^n \int D_\kl(M_{+j}(\cdot|z^{i-1})\|M_{-j}(\cdot|z^{i-1}))\text{d}M_{+j}(z^{i-1}),
\end{equation}
where $M_{+j}(\cdot|z^{i-1})$ and $M_{-j}(\cdot|z^{i-1})$ are the output distributions of $\sK_i$, given the outputs of previous mechanisms $Z^{i-1}=z^{i-1}$, where $X_i$ is distributed according to $P_{+j}$ and $P_{-j}$, respectively. According to Theorem~\ref{Thm:Chi_TV}, we have for any $z^{i-1}$
\begin{equation}
    D_\kl(M_{+j}(\cdot|z^{i-1})\|M_{-j}(\cdot|z^{i-1}))\leq 4\Psi_\eps \tv^2(P_{+j}, P_{-j}), 
\end{equation}
where $\Psi_\eps \coloneqq  e^{-\eps}(e^\eps-1)^2$. Hence, we obtain 
\begin{equation}\label{tv_proof2}
    D_\kl(M^n_{+j}\|M^n_{-j}) \leq  4n \Psi_\eps \tv^2(P_{+j}, P_{-j}).
\end{equation}

Combined with~\eqref{equ:assouad_proof}, we have
\newhz{\[
  R^*(n, \Theta, \ell, \eps) \ge \frac12 k\tau \left[1-  \left( \frac{2n \Psi_\eps}{k}  \cdot { \sum_{j=1}^k \tv^2(P_{+j}\|P_{-j})}
  \right)^{\frac12} \right].
\]}
\subsection{Proof of Corollary~\ref{cor_probability_estimation}}
Let $r \le d$ be an even number which will be specified later. Let $\V = \{-1, +1\}^{r/2}$ and define for a given $\delta\in [0, 1]$, 
$$\theta_v\coloneqq \frac{1}{r} \mathrm{I}_{r} + \frac{\delta}{ r}[v, -v]\in \Delta_r.$$
For any $v \in V$, we let $P^n_v$ be an i.i.d.\ multinomial distribution  with parameter $\theta_v$. Furthermore, we define for any $j\in [r/2]$, 
\begin{equation}
    P^n_{+j} \coloneqq \frac{1}{2^{r/2-1}}\sum_{v:v_j = +1} P^n_v ~~~\text{and}~~~~P^n_{-j} \coloneqq \frac{1}{2^{r/2-1}}\sum_{v:v_j = -1} P^n_v,
\end{equation}
For any $u,v\in V$, it can be verified that for any $p\geq 1$
\begin{equation}
    \|\theta_u - \theta_v\|_h^h \geq 2\cdot\left(\frac{2\delta}{r}\right)^h \left[\sum_{j=1}^{r/2}1_{\{\hat u_j\neq v_j\}}\right].
\end{equation}
Notice that for any $j\in [r/2]$,
\begin{equation}\label{tv_proof3}
    \tv^2(P_{+j}, P_{-j}) = \frac{\delta^2}{r^2}.
\end{equation}
Consequently, Theorem~\ref{thm:private_assouad} implies 
\begin{align}
  R^*(n, \Delta_r, \|\cdot\|^h_h, \eps) & \geq \left(\frac{2\delta}{r}\right)^h \cdot \frac{r}{2} \cdot \left[1-\frac{\delta}{ r}\sqrt{2n\Psi_\eps}\right]. \label{assoud1}
\end{align}
Setting $\delta = \frac{h}{h+1} \cdot \frac{r}{\sqrt{2n\Psi_\eps}}$, we obtain 
\begin{equation}
  R^*(n,\Delta_r, \|\cdot\|_h, \eps)\ge 2^{\frac{1}{2}-\frac1h} \cdot {\frac{h}{h+1}} \cdot \Paren{\frac1{h+1}}^{\frac1h} \cdot  \frac{ r^{1/h}}{\sqrt{n \Psi_\eps}}.
\end{equation}
This bound holds for any $r \le d$ with $\delta \le 1$, hence by choosing $ r = \min \left(d,  \lfloor\frac{h+1}{h}\sqrt{2n\Psi_\eps}\rfloor\right)$, we obtain
\begin{equation}
  R^*(n,\Delta_d, \|\cdot\|_h, \eps)\ge  \min\left\{1, {\frac{\sqrt{2}\cdot h}{h+1}} \cdot \Paren{\frac1{2h+2}}^{\frac1h} \cdot  \frac{d^{1/h}}{\sqrt{n\Psi_\eps}}, {\frac{\sqrt{2}\cdot h}{h+1}} \cdot \Paren{\frac1{\sqrt{2}h}}^{\frac1h} \left[\frac{1}{\sqrt{n \Psi_\eps}}\right]^{1-1/h}\right\}.
\end{equation}

\subsection{Proof of Corollary~\ref{cor:DensityEstimation}}\label{appen:density}
As mentioned in the main body, this corollary can be proved by incorporating Theorem~\ref{thm:private_assouad} into the classical technique of reduction of the density estimation over $\H^\beta_L([0, 1])$ to a parametric estimation problem over a hypercube of a suitable
dimension. For the latter part, we follow the proof of Proposition 2.1 in~\cite{LDP_density_estimation}. 

We begin by describing a standard framework for defining local packing of density functions in $\H^\beta_L([0,1])$. Let $g:\R\to \R$ be an odd function in  $\H^\beta_L([0,1])$ such that $g(x) = 0 $ for any $x\notin [0,1]$. We assume that $g$ satisfies $\|g\|_1<\infty$ which implies that $\|g\|^q_q<\infty$ for any $q>1$. Examples of such function are given in Fig 8 in~\cite{Duchi2016MinimaxOP}.  
Given this function, we define
$$g^b_k(x) \coloneqq 2^{b/2} g (2^b x - k),$$
for some constant $b\geq 0$ (to be determined later) and integers $k\in [N]$, where $N = 2^b-1$. Also, define 
$$f_\theta (x)\coloneqq 1 +\gamma \sum_{k\in [N]} \theta_k g^b_k(x),$$
for $\theta \in [0, 1]^{N}$ and a constant $\gamma$.  Let $\F$ be the collection of all such functions. If $\gamma 2^{b/2}\|g\|_\infty\leq 1$, then $f_\theta\geq 0$ for all $\theta$. Since $g$ is an odd function, we have $\int f_\theta(x)\text{d}x = 1$ for all $\theta$, and thus $f_\theta$ is a density function.  Note also that for any $x, y\in \R$
\begin{align*}
    |f_\theta(x) - f_\theta(y)| & = \gamma \Big|\sum_{k\in [N]}\theta_k\big( g^b_k(x)-g^b_k(y)\big)\Big|\\
    & \leq \gamma \sum_{k\in [N]} \theta_k\big|g^b_k(x)-g^b_k(y)\big|\\
    & = \gamma 2^{b/2} \sum_{k\in [N]} \theta_k\big|g(2^bx-k)-g(2^by-k)\big|\\
    & \leq \gamma 2^{b(\beta + 1/2)} L |x-y|^\beta.
\end{align*}
Thus, if $\gamma 2^{b(\beta + 1/2)}\leq 1$ then $f_\theta\in \H^\beta_L([0,1])$ for all $\theta \in [0, 1]^{N}$, i.e., $\F\subset \H^\beta_L([0,1])$.
Note that for any estimator $\tilde f$ of the density $f$ 
\begin{equation}\label{proof_density1}
    \sup_{f\in\H^\beta_L([0,1])} \E_f[\|\tilde f - f\|^q_q]\geq \sup_{f\in\F} \E_f[\|\tilde f - f\|^q_q] = \max_{\theta\in [0, 1]^N} \E_\theta[\|\tilde f - f_\theta\|^q_q],
\end{equation} 
where $\E_\theta[\cdot]$ denotes the expectation with respect to $f_\theta$. 
Next, we proceed with lower bounding the  $\E_\theta[\|\tilde f - f_\theta\|^q_q]$ for any $\theta\in [0, 1]^N$, as follows 
\begin{align*}
    \E_\theta[\|\tilde f - f_\theta\|^q_q] & = \E_\theta \left[\int |\tilde f(x) - f_\theta(x)|^q\right]\text{d}x\\
    & \geq  \E_\theta \left[\sum_{k=1}^N\int_{k2^{-b}}^{(k+1)2^{-b}} |\tilde f(x) - f_\theta(x)|^q\right]\text{d}x \\
    & =  \sum_{k=1}^N \E_\theta \left[\int_{k2^{-b}}^{(k+1)2^{-b}} |\tilde f(x) - f_\theta(x)|^q\right]\text{d}x\\
    & =  \sum_{k=1}^N \E_\theta \left[\int_{k2^{-b}}^{(k+1)2^{-b}} |\tilde f(x) - \gamma \theta_kg_k^b(x)|^q\right]\text{d}x.
\end{align*}
For each $k\in [N]$, define $\hat f^b_k(x) = \tilde f(x) - \gamma \theta_kg_k^b(x)$ and   
$$\check \theta_k = \argmin_{\theta\in \{0,1\}}\int_{k2^{-b}}^{(k+1)2^{-b}} |\hat f^b_k(x)|^q\text{d}x.$$
Then, according to the Minkowski's inequality, we can write 
\begin{align*}
    2 \left[\int_{k2^{-b}}^{(k+1)2^{-b}} |\hat f^b_k(x)|^q \text{d}x\right]^{1/q}& \geq \left[\int_{k2^{-b}}^{(k+1)2^{-b}} |\tilde f(x) - \gamma \check \theta_kg_k^b(x)|^q\text{d}x\right]^{1/q} + \left[\int_{k2^{-b}}^{(k+1)2^{-b}} |\hat f^b_k(x)|^q\text{d}x\right]^{1/q}\\
    & \geq \left[\int_{k2^{-b}}^{(k+1)2^{-b}} |\gamma \check \theta_kg_k^b(x)-\gamma \theta_kg_k^b(x)|^q \text{d}x\right]^{1/q} \\
    & =  \gamma |\check \theta_k-\theta_k|\left[\int_{k2^{-b}}^{(k+1)2^{-b}} |g_k^b(x)|^q\text{d}x\right]^{1/q},  
\end{align*}
implying 
$$\int_{k2^{-b}}^{(k+1)2^{-b}} |\hat f^b_k(x)|^q \text{d}x\geq \frac{\gamma^q}{2^q} |\check \theta_k-\theta_k|\int_{k2^{-b}}^{(k+1)2^{-b}} |g_k^b(x)|^q\text{d}x = \frac{\gamma^q}{2^q} 2^{\frac{b}{2}(q-2)} |\check \theta_k-\theta_k| \|g\|_q^q .$$
Thus, we have 
\begin{align*}
    \E_\theta[\|\tilde f - f_\theta\|^q_q] &\geq  \frac{\gamma^q}{2^q} 2^{\frac{b}{2}(q-2)}\|g\|_q^q \sum_{k\in [N]}\E_\theta\left[|\check \theta_k-\theta_k|\right] \\
    & =  \frac{\gamma^q}{2^q} 2^{\frac{b}{2}(q-2)}\|g\|_q^q \E_\theta\left[d_H(\check \theta,\theta)\right],
\end{align*}
where $\check \theta = (\check\theta_1, \dots, \check\theta_N)$ and $d_H$ denotes the Hamming distance. Plugging the above into \eqref{proof_density1}, we therefore obtain 
\begin{equation}\label{LB_density}
    \inf_{\tilde f}\sup_{f\in\H^\beta_L([0,1])} \E_f[\|\tilde f - f\|^q_q]\geq \frac{\gamma^q}{2^q} 2^{\frac{b}{2}(q-2)}\|g\|_q^q ~\inf_{\tilde \theta\in [0, 1]^N}\sup_{\theta\in [0, 1]^N}\E_\theta\left[d_H(\tilde \theta,\theta)\right].
\end{equation}
We now suppose $X^n$ are i.i.d.\ samples from either $f_\theta$ or $f_{\theta'}$ with $d_H(\theta, \theta') = 1$. Let $P_\theta$ and $P_{\theta'}$ be the corresponding distributions according to $f_\theta$ and $f_{\theta'}$, respectively. Let $Z_1, \dots, Z_n$ be the outputs of the sequentially interactive mechanism $\sK^n = \sK_1, \dots, \sK_n$ denote obtained from applying $n$ mechanisms $\sK_1, \dots, \sK_n$ (each from $\Q_\eps$) to $X^n$. We denote by $P_\theta^{\otimes n}\sK^n$ the distribution of $Z^n$ when $X^n\sim P_\theta^{\otimes n}$. In this setting, $\tilde \theta$ is an estimator of $\theta$ given $Z^n$.  
Invoking Theorem~\ref{thm:private_assouad} for Hamming distance (with $\tau = \frac{1}{2}$, $k = N$, $P_{+j} = P_\theta$, and  $P_{-j} = P_{\theta'}$), we thus obtain 
\begin{equation}\label{tsybakov_31}
\inf_{\sK_1, \dots, \sK_n\in \Q_\eps}\inf_{\tilde \theta\in [0, 1]^N}\sup_{\theta\in [0, 1]^N}\E_\theta\left[d_H(\tilde \theta,\theta)\right]\geq \frac{N}{4} \left[1-\sqrt{2n\Psi_\eps\tv^2(P_\theta, P_{\theta'})}\right].
\end{equation}
Since $d_H(\theta, \theta')=1$, we can bound $\tv(P_\theta, P_{\theta'})$ as follows
\begin{align*}
    \tv(P_\theta, P_{\theta'}) &= \frac{1}{2}\int\left|f_\theta(x)- f_{\theta'}(x)\right|\text{d}x \\
    & = \frac{\gamma}{2} \int\Big|\sum_{k\in [N]}(\theta_k-\theta'_k)g_k^b(x)\Big|\text{d}x\\
    & = \frac{\gamma}{2} 2^{-b/2} \|g\|_1.
\end{align*}
Thus, we obtain 
\newhz{\begin{equation}\label{tsybakov_3}
\inf_{\tilde \theta\in [0, 1]^N}\sup_{\theta\in [0, 1]^N}\E_\theta\left[d_H(\tilde \theta,\theta)\right]\geq \frac{N}{4} \left[1-\sqrt{0.5n\Psi_\eps\gamma^22^{-b}\|g\|_1^2}\right].
\end{equation}}
Let 
\begin{equation}\label{choice_N}
    \gamma = (n\Psi_\eps)^{-\frac{2\beta +1}{2(2\beta +2)}}, ~~~\text{and}~~~ N = (n\Psi_\eps)^{\frac{1}{2\beta +2}}.
\end{equation}
It can be verified that for these choices of $\gamma$ and $N$ (or equivalently $b$), we have $n\Psi_\eps\gamma^22^{-b}\leq  1$ (note also that both previous assumptions  $\gamma 2^{b(\beta+1/2)}\leq 1$ and $\gamma 2^{b/2}\lesssim 1$ are now satisfied.) Thus, we deduce
\begin{equation}\label{tsybakov_4}
\inf_{\tilde \theta\in [0, 1]^N}\sup_{\theta\in [0, 1]^N}\E_\theta\left[d_H(\tilde \theta,\theta)\right]\gtrsim (n\Psi_\eps)^{\frac{1}{2\beta +2}}.
\end{equation}
Finally, in light of \eqref{LB_density}, we can write 
$$\sup_{f\in\H^\beta_L([0,1])} \E_f[\|\tilde f - f\|^q_q]\gtrsim (n\Psi_\eps)^{-\frac{q\beta}{2\beta+2}}.$$

\subsection{Proof of Theorem~\ref{thm:MIM}}\label{appen:MIM}
Fix $n$ mechanisms $\sK_1, \dots, \sK_n$ each of which is $\eps$-LDP. 
Let $\pi$ be the distribution of $\theta$. Given a realization of $\theta$, we sample $n$ i.i.d.\ samples $X^n$ from $P_\theta$, and thus $P_X(A) = \int P_\theta(A)\text{d}\pi$ for $A\subset \X$.
Notice that for any estimate $\hat\theta = \psi(Z^n)$ of $\theta$ \newhz{that achieves $R^*_\pi(n, \|\cdot\|^r, \eps)$}, we have 
\begin{equation}\label{SLB1}
    D^*_\pi(n, \Theta, \|\cdot\|^r, \eps)\coloneqq \inf_{\substack{P_{\hat\theta|\theta}:\\\E_\pi[\|\theta-\hat\theta\|^r]\leq R^*_\pi(n, \|\cdot\|^r, \eps)}}I(\theta; \hat\theta)\leq I(\theta; \hat\theta). 
\end{equation}
In the following, we obtain a lower bound for $D^*_\pi(n, \Theta, \|\cdot\|^r, \eps)$ and an upper bound for $I(\theta; \hat\theta)$. 

We first discuss how to lower bound $D^*_\pi(n, \Theta, \|\cdot\|^r, \eps)$. Invoking Shannon Lower Bound (see e.g., \cite{ShannonLowerBound} or Problem 10.6 in~\cite{cover2012elements}), we can write
\begin{equation}\label{SLB2}
    D^*_\pi(n, \Theta, \|\cdot\|^r, \eps)\geq H(\theta) - \frac{d}{r}\log\left[\frac{re R^*_\pi}{d}\left(V_d \Gamma(1+d/r)\right)^{r/d}\right],
\end{equation}
where $R^*_\pi\coloneqq R^*_\pi(n, \Theta, \|\cdot\|^r, \eps)$. 

Now, we derive an upper bound for $I(\theta, \hat\theta)$. First, notice that the data processing inequality implies
$$I(\theta; \hat\theta)\leq I(\theta; Z^n).$$ 
We now seek to derive an upper bound for $ I(\theta; Z^n)$. To this goal, we rely on the distribution-dependent version of \eqref{eq:Contraction_MI} to connect the decay of mutual information over the Markov chain $\theta-X^n-Z^n$ with $\eta_\kl(P^{\otimes n}_X, \sK^n)$, where $P^{\otimes n}_X$ is the product distribution corresponding to $P_X$ and $\sK^n$ denotes the sequentially interactive mechanism obtained from $n$ mechanisms $\sK_1, \dots, \sK_n$.  In fact, it can be shown that (see \cite{Anantharam_SDPI} and Appendix B in~\cite{Yury_Dissipation} for a proof in the discrete and general cases, respectively)
\begin{equation}\label{eq:Contraction_MI_inputdependent}
    \eta_\kl(P^{\otimes n}_X, \sK^n) = \sup_{\substack{P_{U|X^n}:\\ U-X^n-Z^n}}~\frac{I(U; Z^n)}{I(U; X^n)},
\end{equation}
for any channel $P_{U|X^n}$ satisfying the Markov chain $U-X^n-Z^n$. Therefore, we can write 
\begin{align}
    I(\theta; Z^n)&\leq I(\theta; X^n) \eta_\kl(P^{\otimes n}_X, \sK^n)\nonumber\\
    &\leq I(\theta; X^n)\max_{i\in [n]}~\eta_\kl(P_X, \sK_i)\nonumber\\
    &\leq  I(\theta; X^n) \max_{i\in [n]}~\eta_\kl(\sK_i) \nonumber\\
    &\leq \Upsilon_\eps I(\theta; X^n), \label{SDPI_MI2}
\end{align}
where the first step follows from \eqref{eq:Contraction_MI_inputdependent}, the second steps is due to the tensorization of the distribution-dependent contraction coefficient under kl-divergence (see Appendix~\ref{Appen:tensotization}), and the last step is an application of Theorem~\ref{thm_chiSDPI}.

Plugging the lower bound \eqref{SLB2} and the upper bound \eqref{SDPI_MI2} into \eqref{SLB1}, we obtain
\begin{equation}\label{SLB_UB1}
    H(\theta) - \frac{d}{r}\log\left[\frac{re R^*_\pi}{d}\left(V_d \Gamma(1+d/r)\right)^{r/d}\right]\leq \Upsilon_\eps I(\theta; X^n),
\end{equation}
from which the desired result follows. 

\subsection{Proof of Corollary~\ref{cor:GaussianLocation_general}}\label{appen:GaussianLocation}
Let $\pi$ be uniform distribution on $\tilde \Theta\subset \Theta$ and $\theta\sim \pi$. Given any realization of $\theta$, we pick $n$ i.i.d.\ samples $X^n$ from $\N(\theta, \sigma^2 \mathrm{I}_d)$.  It can be shown that $\bar X_n\coloneqq \frac{1}{n}\sum_{i=1}^n X_i$ is a sufficient statistics for $\theta$ and hence 
\begin{align}
    I(\theta; X^n) &= I(\theta; \bar X_n) \nonumber\\
    & \leq \inf_{Q}\sup_{\theta\in \tilde\Theta} D_\kl\Big(\N\big(\theta, \frac{\sigma^2}{n}\mathrm{I}_d\big)\Big\|
    Q\Big)\nonumber\\
      & \leq \inf_{\theta'\in \tilde\Theta}\sup_{\theta\in \tilde\Theta} D_\kl\Big(\N\big(\theta, \frac{\sigma^2}{n}\mathrm{I}_d\big)\Big\|
    \N\big(\theta', \frac{\sigma^2}{n}\mathrm{I}_d\big)\Big)\nonumber\\
    & = \inf_{\theta'\in \tilde\Theta}\sup_{\theta\in \tilde\Theta} \frac{n}{2\sigma^2}\|\theta-\theta'\|_2^2\nonumber\\
    &= \frac{n}{2\sigma^2}\mathsf{rad}(\tilde\Theta)^2, 
\end{align}
where $\mathsf{rad}(\Theta)$ denotes the $\ell_2$-radius of $\Theta$. 
Plugging this upper bound and $H(\theta) = \log V(\tilde \Theta)$ in Theorem~\ref{thm:MIM} (or in \eqref{SLB_UB1} more specifically), we obtain   
\begin{equation}
\log V(\tilde\Theta) - \frac{d}{r}\log\left[\frac{re R^*_\pi}{d}\left(V_d \Gamma(1+d/r)\right)^{r/d}\right]\leq \frac{n}{2\sigma^2}\mathsf{rad}(\tilde\Theta)^2, \end{equation}
which, after a re-arrangement, leads to 
\begin{align}
    R^*_\pi&\geq \frac{d}{re [V_d\Gamma(1+d/r)]^{r/d}}\left[V(\tilde\Theta)e^{-\frac{n}{2\sigma^2}\mathsf{rad}(\tilde\Theta)^2\Upsilon_\eps}\right]^{r/d}\nonumber\\
    & = \frac{d}{re [V_d\Gamma(1+d/r)]^{r/d}}\left[\frac{V(\tilde\Theta)}{V_2(\tilde \Theta)}\right]^{r/d}\left[V_2(\tilde \Theta)e^{-\frac{n}{2\sigma^2}\mathsf{rad}(\tilde\Theta)^2\Upsilon_\eps}\right]^{r/d} ,\label{proof1}
\end{align}
where $V_2(\tilde \Theta)$ is the volume of the $\ell_2$-ball of the same radius as $\tilde\Theta$. 
We now maximize the right hand-side in \eqref{proof1} over the choice of $\tilde \Theta$. To this end, first note that 
\begin{equation}
     \sup_{\tilde\Theta\subseteq \Theta}\left[\frac{V(\tilde\Theta)}{V_2(\tilde \Theta)}\right]^{r/d}\left[V_2(\tilde \Theta)e^{-\frac{n}{2\sigma^2}\mathsf{rad}(\tilde\Theta)^2\Upsilon_\eps}\right]^{r/d} \geq \left[\frac{V(\Theta)}{V_2(\Theta)}\right]^{r/d}\sup_{\tilde\Theta\subseteq \Theta}\left[V_2(\tilde \Theta)e^{-\frac{n}{2\sigma^2}\mathsf{rad}(\tilde\Theta)^2\Upsilon_\eps}\right]^{r/d}.\label{proof11}
\end{equation}
Recall that 
\begin{equation}
    V_2(\tilde \Theta) = \frac{\pi^{d/2}\mathsf{rad}(\tilde\Theta)^d}{\Gamma(1+d/2)}.
\end{equation}
Hence, the maximization in \eqref{proof11} can be written as 
\begin{align}
     \sup_{\tilde\Theta\subseteq \Theta}\left[V_2(\tilde \Theta)e^{-\frac{n}{2\sigma^2}\mathsf{rad}(\tilde\Theta)^2\Upsilon_\eps}\right]^{r/d} &  = \frac{\pi^{r/2}}{\Gamma(1+d/2)^{r/d}} \sup_{\tilde\Theta\subseteq \Theta}\left[\mathsf{rad}(\tilde\Theta)^2 e^{-\frac{n}{\sigma^2 d}\mathsf{rad}(\tilde\Theta)^2\Upsilon_\eps}\right]^{r/2}\nonumber\nonumber\\
    & = \frac{\pi^{r/2}}{\Gamma(1+d/2)^{r/d}} \left[\frac{\sigma^2 d}{n\Upsilon_\eps}\right]^{r/2}\sup_{x\leq \frac{n}{\sigma^2d}\mathsf{rad}(\theta)^2\Upsilon_\eps}\left[xe^{-x}\right]^{r/2}\nonumber\\
    & \geq   \frac{\pi^{r/2}}{e\Gamma(1+d/2)^{r/d}} \min\left\{\mathsf{rad}(\Theta)^r, \Big[\frac{\sigma^2 d}{n\Upsilon_\eps}\Big]^{r/2}\right\}\label{proof2}\\
    & \geq \frac{1}{e}\left[\frac{\pi}{ 2d}\right]^{r/2} \min\left\{\mathsf{rad}(\Theta)^r, \Big[\frac{\sigma^2 d}{n\Upsilon_\eps}\Big]^{r/2}\right\}\label{proof22}\\
    & \geq \frac{1}{ed^{r/2}} \min\left\{\mathsf{rad}(\Theta)^r, \Big[\frac{\sigma^2 d}{n\Upsilon_\eps}\Big]^{r/2}\right\}.\label{proof33}
\end{align}
where \eqref{proof2} follows from the fact that $x\mapsto xe^{-x}$ is increasing over $[0, 1]$ and decreasing over $[1, \infty)$ and thus it attains its global maximum of $\frac{1}{e}$ at $x=1$. Also, \eqref{proof22} is due to the fact that $\Gamma(1+d/2) \leq \frac{d}{2} (\frac{d}{2})^{d/2}$, thus $(\Gamma(1+d/2))^{1/d} \leq \sqrt{d/2} (\frac{d}{2})^{1/d}\leq \sqrt{2d}$.
Plugging \eqref{proof33} and \eqref{proof11} into  \eqref{proof1}, we obtain the desired result.

\subsection{Proof of Lemma~\ref{lemma:BHT_SampleComplexity}}\label{appen:BHT}

Recall that $Z^n$ is the output of the sequentially interactive mechanisms $\sK_1, \dots, \sK_n$ each of which is $\eps$-LDP. Let $M_0^n$ and $M_1^n$ be the distribution of $Z^n$ under the null and alternative hypothesis, respectively. Let $N_i^n$ be the distribution induced on $Z^n$ when $X_i\sim Q$ for $i\in [n]$ and the remaining $X_j\sim P$ for $j\in [n]/\{i\}$. We wish to obtain an upper bound on $H^2(M_0^n, M_1^n)$. To do so, we use \cite[Lemma 2]{Communication_Complexity_SDPI} (see also \cite[Lemma 11.7]{Duchi_LectureNote}) that states 
\begin{equation}\label{Proof_YBHellinger1}
    H^2(M_0^n, M_1^n) \leq 7 \sum_{i=1}^n H^2(M_0^n, N_i^n).
\end{equation}
An upper bound on $H^2(M_0^n, N_i^n)$ for an arbitrary $i\in [n]$ thus implies an upper bound on $H^2(M_0^n, M_1^n)$. To derive an upper bound on $H^2(M_0^n, N_i^n)$, we first notice that $M_0^n(z^n) = \prod_{j=1}^nM_0^j(z_j|z^{j-1})$ and $N_i^n(z^n) = \prod_{j=1}^nN_i^j(z_j|z^{j-1})$ for any $z^n\in \Z^n$, where $M_0^j$ and $N_i^j$ are the conditional distributions of $Z_j$ conditioned on $Z^{j-1}$  corresponding to $M_0^n$ and $N_i^n$, respectively.  Then, we can write
\begin{align}
    H^2(M_0^n, N_i^n) &\leq \frac{1}{2} D_\kl(M_0^n\| N_i^n)\nonumber\\
    &= \frac{1}{2} \sum_{j=1}^n \int D_\kl \left(M_0^{j} (\cdot|z^{j-1}) \| N_i^{j} (\cdot|z^{j-1}) \right)\text{d} M_0^{j-1}(z^{j-1}), \label{equ:chain_rule}
\end{align}
where the inequality is due to Pinsker's inequality, and the equality follows from the chain rule for KL divergence. 
Note that $M_0^j (\cdot|z^{j-1}) = N_i^j (\cdot|z^{j-1})$ except for $j=i$. Therefore, we have  
\begin{align}
    H^2(M_0^n, N_i^n) & \leq \frac{1}{2} \int D_\kl \left(M_0^{i} (\cdot|z^{i-1}) \| N_i^{i} (\cdot|z^{i-1}) \right)\text{d} M_0^{i-1}(z^{i-1})\nonumber\\
    & \leq \frac{1}{2} \chi^2(P\sK_i\| Q\sK_i)\nonumber\\
    & \leq 2\Psi_\eps\tv^2(P, Q),\label{equ:chain_rule2}
\end{align}
where the second inequality follows from Theorem~\ref{Thm:Chi_TV}.

We can analogously apply the tensorization property of the squared Hellinger distance (see, e.g., \cite[Remark 7.7]{polyanskiy_book}) in the same fashion as \eqref{equ:chain_rule} to obtain
\begin{align}
    H^2(M_0^n, N_i^n) & \leq H^2(P\sK_i, Q\sK_i)\nonumber\\
    & \leq \Upsilon_\eps H^2(P, Q),\label{equ:chain_rule3}
\end{align}
where the second inequality is due to Theorem~\ref{thm_chiSDPI}.


Plugging \eqref{equ:chain_rule2} and \eqref{equ:chain_rule3} into \eqref{Proof_YBHellinger1}, we arrive at 
\begin{equation}\label{eqn:LB_Hellinger}
    H^2(M_0^n, M_1^n) \leq 7  n \min\{\Upsilon_\eps H^2(P, Q), 2\Psi_\eps\tv^2(P, Q)\}.
\end{equation}

On the other hand, since  $$H^2(M_0^n, M_1^n)  \geq 2-2\sqrt{1-\tv^2(M_0^n, M_1^n)},$$ and both error probabilities are smaller than $\frac{1}{10}$, it follows that $H^2(M_0^n, M_1^n) \geq \frac{4}{5}$. 
This, together with \eqref{eqn:LB_Hellinger}, implies that 
$$\mathsf{SC}^{P, Q}_\eps \geq \frac{4}{35}\max\Big\{\frac{1}{\Upsilon_\eps H^2(P, Q)}, \frac{1}{2\Psi_\eps\tv^2(P, Q)}\Big\}.$$

To prove the upper bound, we consider the following simple setting: $\sK_1, \dots, \sK_n$ are non-interactive mechanisms and all are of the same form, that is, $\sK_i = \sK$ for some $\sK\in \Q_\eps$. In this case, $Z_1, \dots, Z_n$ are i.i.d. samples distributed according to $P\sK$ and $Q\sK$ under the null and alternative hypotheses,  respectively. Therefore, in light of \cite[Theorem 2]{folklore_BHT_sample}, we can write 
$$ \mathsf{SC}^{P, Q}_\eps \leq \frac{2\log(5)}{H^2(P\sK, Q\sK)},$$
for any choice of $\sK\in \Q_\eps$. Since $\tv^2(P, Q)\le H^2(P, Q)$ for any distributions $P$ and  $Q$, it follows that  
$$ \mathsf{SC}^{P, Q}_\eps \leq \frac{2\log(5)}{\tv^2(P\sK, Q\sK)},$$
for any choice of $\sK\in \Q_\eps$. Now, we let $\sK$ be the binary mechanism, defined in \eqref{binary_mech}. For this mechanism, we know that \cite{kairouz2014extremal_JMLR}
    \begin{equation}
        \tv(P\sK\|Q\sK) = \frac{e^\eps-1}{e^\eps+1}\tv(P, Q). 
    \end{equation}
Thus, we obtain
    \begin{equation}
        \mathsf{SC}^{P, Q}_\eps \le   \frac{2\log (5)}{\Upsilon_\eps\tv^2(P, Q)}.
    \end{equation}

\section{Private  Distribution Estimation -- Upper Bound}\label{appen:hadamard}
 In this section, we continue our discussion on the locally private distribution estimation problem in Section~\ref{sec:LeCam}. Recall that in Corollary~\ref{cor_probability_estimation} we showed that 
For any $h\geq 1$ and $\eps\geq 0$, we have 
\begin{align*}
    R^*(n,\Delta_d, \|\cdot\|_h, \eps)&\ge  \min\bigg\{1, {\frac{\sqrt{2}h}{h+1}} \Big[\frac1{2h+2}\Big]^{1/h} \frac{d^{1/h}}{\sqrt{n\Psi_\eps}},{\frac{\sqrt{2}h}{h+1}} \Big[\frac1{\sqrt{2}h}\Big]^{1/h} \Big[\frac{1}{\sqrt{n \Psi_\eps}}\Big]^{1-1/h}\bigg\}.
\end{align*}    
Here, we seek to derive an upper bound for $R^*(n,\Delta_d, \|\cdot\|_h, \eps)$.

\begin{theorem}\label{thm:hadamard}
    For any $2\le h\leq 100$ and $\eps\geq 0$, when $n \gtrsim {\min\Paren{d^{\frac{2}{h}}, \Paren{e^\eps}^{\frac{2}{h}}}}$, we have 
    $$R^*(n, \Delta_d,\|\cdot\|_h, \eps) \lesssim 
 {\frac{\Paren{e^\eps}^{\frac{h-1}{h}} \cdot \Paren{e^\eps+d}^{\frac1h}} { (e^\eps-1)\sqrt{n}} }.
$$
\end{theorem}

\begin{proof}
We adopt the same algorithm as the one described in Section 5 in~\cite{Disribution_estimation_hadamard}, and generalize their analysis to any $h\ge 2$. We also follow their notation as well.

Let $\hat{p}(x)$ be our estimate of $p(x)$. By Equation (22), Appendix C in~\cite{Disribution_estimation_hadamard}, we have 
\begin{align*}
    \hat{p}(x) - p(x) = \frac{2(2B-1+e^\eps)}{e^\eps-1}\cdot \Paren{\Paren{\widehat{p(C_x)}-p(C_x)} - \frac12 \Paren{\widehat{p(S_i)} - p(S_i)}}.
\end{align*}
Note that for any $a,b\in \mathbb{R}$, and $h\ge 1$, $\absv{a-b}^h \le 2^h \Paren{\absv{a}^h + \absv{b}^h}$. Therefore,
\begin{align*}
    \expectation{\absv{\hat{p}(x) - p(x)}^h} = \frac{2^{2h}\cdot (2B-1+e^\eps)^h}{(e^\eps-1)^h}\cdot \Paren{ \expectation{\absv{\widehat{p(C_x)}-p(C_x)}^h }+ \frac1{2^h} \expectation{ \absv{\widehat{p(S_i)} - p(S_i)}^h }}.
\end{align*}
We now to proceed to upper bound both terms inside the parenthesis in the right-hand side of the above identity. To do so, we need the following lemma which was first proved by Steinke  \cite{SteinkeTwitter}. 

\begin{lemma}[\cite{SteinkeTwitter}]
\label{lem:thomas}
Given a random variable $Z$ drawn from a binomial distribution $\binomial(n,p)$, i.e., $Z \sim \binomial(n,p)$, for any $2 \le h \le 100$, there exists a universal constant $c_2$ such that 
\begin{align*}
    \expectation{\absv{Z-np}^h} \le c_2 \cdot \max \Paren{1, (np)^{\frac{h}{2}}},
\end{align*}
\end{lemma}
\begin{proof}[Proof of Lemma~\ref{lem:thomas}]
Let $Z \sim \binomial(n,p)$. We first consider the case when $h\ge2$ is an even integer. Then for all $t \in \mathbb{R}$,
\begin{align*}
    \expectation{e^{tZ}} = (1-p+p\cdot e^t)^n \le \Paren{e^{p\cdot(e^t-1)}}^n.
\end{align*}
Note that $h \ge 2$ is an even integer. By the Taylor expansion of the exponential function, 
\begin{align*}
    1+ \frac{t^h}{h!}\cdot \expectation{(Z-np)^h} &\le \frac{\expectation{e^{t(Z-np)}}+\expectation{e^{-t(Z-np)}}}{2} \\
    & \le \frac{e^{np(e^t-1-t)}+e^{np(e^{-t}-1+t)} }{2} \le e^{np(e^t-1-t)},
\end{align*}
for all $t \ge 0$.
Thus,
$$  \expectation{(Z-np)^h} \le \inf_{t\ge 0} h! \cdot t^{-h} \Paren{e^{np(e^t-1-t)}-1}.$$
By setting $t = \min\Paren{1,\sqrt{h/np}}$ so that $e^t-1-t\le t^2 = \min\Paren{1, h/np}$,
\begin{align*}
\expectation{(Z-np)^h} &\le h! \cdot \max\Paren{1,(h/np)^{-h/2}} \cdot e^{h} \\
&\le c_1\Paren{h^{h+\frac12}\cdot \max\Paren{1, (np)^{\frac{h}{2}}}},
\end{align*}
where the last inequality comes from Stirling's approximation, and $c_1$ is a universal constant.

Now we generalize our analysis to the case when $h \in \mathbb{R}$. 
Let $h^\prime \ge 4$ be an even integer. Given $h \in [h^\prime -2, h^\prime)$, by Jensen's inequality,
\begin{align*}
\expectation{\absv{Z-np}^h} &\le \Paren{\expectation{(Z-np)^{h^\prime}}}^{h/h^\prime}
 \\
&\le c_1 (h^{\prime})^{h+\frac12} \cdot \max \Paren{1, (np)^{\frac{h}{2}}}\\
&\le c_2 \max \Paren{1, (np)^{\frac{h}{2}}},
\end{align*}
where the last inequality comes from the fact that $h^\prime /h \le 2$ and $h\le 100$, and $c_2$ is another universal constant.

\end{proof}

By Equations (13) and (14) in~\cite{Disribution_estimation_hadamard}, $n \cdot \widehat{p(C_x)} \sim \binomial(n,p(C_x))$ and $n \cdot \widehat{p(S_i)} \sim \binomial(n,p(S_i))$.
Therefore, by Lemma~\ref{lem:thomas}, we have 
\begin{align*}
    \expectation{\absv{\widehat{p(C_x)}-p(C_x)}^h } \le \frac{c_2}{n^h} + c_2 (n^{-1}\cdot p(C_x))^{\frac{h}{2}},
\end{align*}
and
\begin{align*}
    \expectation{\absv{\widehat{p(S_i)}-p(S_i)}^h } \le \frac{c_2}{n^h} + c_2 (n^{-1}\cdot p(S_i))^{\frac{h}{2}},
\end{align*}

Summing over $x$,
\begin{align*}
    \expectation{\sum_x\absv{\widehat{p(C_x)}-p(C_x)}^h } &\le c_2 \Paren{n^{-\frac{h}{2}} \cdot \sum_x \Paren{p(C_x)}^{\frac{h}{2}} + \frac{ d}{n^h}} \\ &\le  c_2 \Paren{ n^{-\frac{h}{2}} \cdot \Paren{\frac{d}{2B-1+e^\eps}+\frac{e^\eps-1}{2(2B-1+e^\eps)}+ \frac{b(e^\eps-1)}{2(2B-1+e^\eps)}}+\frac{d}{n^h}}\\
    &=  c_2 \Paren{n^{-\frac{h}{2}} \cdot \frac{2d+(e^\eps-1)(b+1)}{2(2B-1+e^\eps)}+\frac{d}{n^h}},
\end{align*}
where the last inequality comes from $\sum_x \Paren{p(C_x)}^{\frac{h}{2}} \le \sum_x p(C_x)$, and Equation (25) in~\cite{Disribution_estimation_hadamard}, i.e.,
$${\sum_x p(C_x)} \le {\frac{d}{2B-1+e^\eps}+\frac{e^\eps-1}{2(2B-1+e^\eps)}+ \frac{b(e^\eps-1)}{2(2B-1+e^\eps)}}.$$

Similarly, by Equation (26) in~\cite{Disribution_estimation_hadamard},
\begin{align*}
    \expectation{\sum_x\absv{\widehat{p(S_i)}-p(S_i)}^h } &\le c_2 \Paren{ n^{-\frac{h}{2}} \Paren{\frac{2d}{2B-1+e^\eps}+ \frac{b(e^\eps-1)}{2(2B-1+e^\eps)}}+\frac{d}{n^h}} \\
    &= c_2\Paren{n^{-\frac{h}{2}} \cdot \frac{b(e^\eps-1)+4d}{2(2B-1+e^\eps)}+\frac{d}{n^h}}.
\end{align*}

Summing up the two terms, we have 

\begin{align*}
\expectation{\sum_x\absv{\hat{p}(x) - p(x)}^h} \le \frac{c_2 \cdot 2^{2h}\cdot (2B-1+e^\eps)^h}{(e^\eps-1)^h}\cdot \Paren{ { n^{-\frac{h}{2}} \cdot \frac{ 6d+(e^\eps-1)(2b+1)}{2(2B-1+e^\eps)}}+\frac{2d}{n^h}}.
\end{align*}

Finally, by the Jensen's inequality,

\begin{align}
\expectation{\Paren{\sum_x\absv{\hat{p}(x) - p(x)}^h}^\frac{1}{h}}&\le \Paren{\expectation{\sum_x\absv{\hat{p}(x) - p(x)}^h}}^{\frac1h}\nonumber\\&
\le \frac{4 \cdot (c_2)^{\frac1h} \cdot (2B-1+e^\eps)}{e^\eps-1} \cdot\Paren{ n^{-\frac12}\Paren{\frac{ 6d+(e^\eps-1)(2b+1)}{2(2B-1+e^\eps)}}^{\frac1h}+ \frac{(2d)^{\frac1h}}{n}}\nonumber \\
&= 4 \cdot \Paren{\frac{c_2}{2}}^{\frac1h} \cdot \frac{\Paren{2B-1+e^\eps}^{\frac{h-1}{h}}\cdot \Paren{6d+(e^\eps-1)(2b+1)}^{\frac1h}}{\sqrt{n}(e^\eps-1)}+\nonumber\\
& ~~~~~\frac{4 \cdot (c_2)^{\frac1h} \cdot (2B-1+e^\eps)}{e^\eps-1}\cdot \frac{(2d)^{\frac1h}}{n}.
\end{align}
Noting that $B=\Theta(\min(e^\eps, 2k))$ and $b=\Theta(k/B+1)$, we have
$$
\expectation{\Paren{\sum_x\absv{\hat{p}(x) - p(x)}^h}^\frac{1}{h}}  \lesssim {\frac{\Paren{e^\eps}^{\frac{h-1}{h}} \cdot \Paren{e^\eps+d}^{\frac1h}} { (e^\eps-1)\sqrt{n}}+ \frac{e^\eps \cdot d^{\frac1h}}{(e^\eps-1)n}}.
$$
Note that the first dominates when $n \gtrsim \Paren{\min\Paren{d^{\frac{2}{h}}, \Paren{e^\eps}^{\frac{2}{h}}}}$.
\end{proof}

\end{document}